\documentclass[aps,prl,reprint,superscriptaddress,nofootinbib,floatfix]{revtex4-2}
\usepackage[OT1]{fontenc}

\usepackage{amsthm,amsmath,amssymb}
\usepackage{mathrsfs}
\usepackage{graphicx}
\usepackage{dcolumn}
\usepackage{bm}
\usepackage{amsfonts}
\usepackage{hyperref}
\renewcommand*{\HyperDestNameFilter}[1]{\jobname-#1}
\hypersetup{hidelinks}
\usepackage{xcolor}
\begin{document}
 \title{Overcoming Backaction-Induced Bias in Quantum Temporal Correlations}

\author{Yi-Mu Du}
\email{ymdu@gscaep.ac.cn}
 \affiliation{Graduate School of China Academy of Engineering Physics, Beijing 100193, China}
\date{October 30, 2026}

\begin{abstract}
Temporal correlations are central to quantum spectroscopy and many-body dynamics. They are typically inferred from sequential measurements, but measurement backaction generally biases the estimates. We ask whether measurements with adjustable strength admit universal unbiased estimators that require no knowledge of the initial state or dynamics. For nested commutator and anticommutator correlations involving an arbitrary number of time points, we prove a no-go theorem for conventional measurement protocol that characterizes the spectral obstructions to the existence of such estimators. To overcome these obstructions, we propose a randomized spectral phase control protocol and jointly optimize its control probabilities, estimators, and measurement strength to achieve universal unbiased estimation while minimizing worst-case variance. A collective-spin model identifies regimes in which the optimized controlled protocol requires fewer samples than the conventional protocol. These results reveal the spectral origin of the limits imposed by measurement backaction on universal unbiased estimation and provide a method to overcome them.
\end{abstract}

\maketitle
\textit{Introduction.}--Temporal correlation functions encode fluctuation spectra and linear and nonlinear response, linking spectroscopic signals to quantum many-body dynamics \cite{kubo1957,degen2017,altman2021,li2016,li2013,altuntas2025,geier2022,rostami2021,mukamel2003,cheung2025,sun2026symmetries,meinel2022}. Experiments typically measure observables sequentially at different times and estimate these correlations from the joint statistics of the outcomes \cite{mitchison2007,diosi2016,pfender2019,wang2019}. However, earlier measurements disturb the system and alter the statistics of later outcomes, so measurement backaction generally biases the resulting correlation estimates.

Existing approaches suppress or correct this bias by exploiting special properties of the system or observables, using additional measurement information \cite{kastner2020,dressel2018,mohseninia2019,delre2024,pengfei2024}, or combining weaker coupling with suitable detector design and data processing \cite{uhrich2017,kastner2018,bednorz2013,bulte2018}. However, methods relying on special properties have limited applicability, while hierarchical reconstruction requires additional measurements of lower-order correlations \cite{delre2024}. Although weaker coupling can make the bias vanish asymptotically, it increases statistical fluctuations and the number of samples needed to achieve a given precision. This raises a basic question: can sequential measurements with adjustable strength provide a universal unbiased estimator of a specified temporal correlation with arbitrary number of time points? Here, universality means that the same estimator remains unbiased for any initial state and free evolution, without depending on either.

We characterize the estimable operator families for sequential Gaussian measurements. A single-shot estimator is a function of one complete measurement record; averaging its values over independent repetitions gives the sample-mean estimator. For nested commutator and anticommutator correlations relevant to quantum noise measurements and nonlinear spectroscopy \cite{mukamel2003,meinel2022,wang2019,wan2019}, we prove that, without additional control, the spectra of the measured observables alone determine whether a universal unbiased single-shot estimator exists. We derive a necessary and sufficient criterion and establish a no-go theorem that identifies a spectral obstruction to universal unbiased estimation. To overcome this obstruction, we use the idea of probabilistic error cancellation (PEC) \cite{Temme2017,Endo2018,HernandezGomez2024} by applying random spectral phase flips between successive measurements and combining the control records with weighted outcomes, achieving universal unbiased estimation at finite measurement strength. Exact bias cancellation does not fix the sampling cost, since different unbiased constructions can produce different fluctuations. The independence of the controlled estimable space from finite nonzero measurement strength provides further freedom to improve statistical precision. We therefore optimize within the family of universally unbiased protocols and use a collective spin model to identify when the controlled protocol offers a sampling advantage over the conventional protocol.

\textit{Unbiased estimation and randomized control.---}
Consider a quantum system in an initial state $\rho_0$.
We measure observables $A_0,\ldots,A_{N-1}$ at times
$t_0<\cdots<t_{N-1}$.
Their spectral decompositions are
$A_j=\sum_{a=1}^{D_j}\lambda_{j,a}P_{j,a}$, where $D_j$ is the number
of distinct eigenvalues and $P_{j,a}$ projects onto the full eigenspace
of $\lambda_{j,a}$.
Between consecutive measurements, the system undergoes free evolution described by
$\mathbf U=(U_{1:0},U_{2:1},\ldots,U_{N-1:N-2})$.

Each measurement couples the system to an independent Gaussian probe
through $W_j=\exp(-i\gamma A_j\otimes p_j)$. Probe position and momentum
are in conjugate units with initial variances $1$ and $1/4$; the common
strength $\gamma>0$ is the coupling divided by the initial position width.
For readout type $\eta_j=x,p$, the recorded value $y_j$ is the probe
position divided by $\gamma$ or the probe momentum, respectively.
The unnormalized conditional system state is
$K_{j,\gamma}^{\eta_j}(y_j)\rho(t_j)
 K_{j,\gamma}^{\eta_j}(y_j)^\dagger$, with Kraus operators
\begin{subequations}\label{eq:kraus}
\begin{align}
K_{j,\gamma}^{x}(y_j)
&=\left(\frac{\gamma^2}{2\pi}\right)^{1/4}
\exp\!\left[-\frac{\gamma^2}{4}(y_j-A_j)^2\right],
\label{eq:kraus-x}\\
K_{j,\gamma}^{p}(y_j)
&=\left(\frac{2}{\pi}\right)^{1/4}
e^{-y_j^2}e^{-i\gamma y_j A_j}.
\label{eq:kraus-p}
\end{align}
\end{subequations}

Without added control, this is the conventional protocol, with record $\mathbf y=(y_0,\ldots,y_{N-1})$.
The joint effect density is
$E_{\gamma,\mathbf U}^{\boldsymbol\eta}(\mathbf y)=B^\dagger B$, where
\[
B=\overleftarrow{\prod}_{j=0}^{N-1}
\bigl(U_{j+1:j}K_{j,\gamma}^{\eta_j}(y_j)\bigr).
\]
The arrow places later operations to the left.
Here $U_{N:N-1}$ is an arbitrary unitary after the final readout;
it cancels from $B^\dagger B$.
The joint probability density is
$p_{\gamma,\rho_0,\mathbf U}^{\boldsymbol\eta}(\mathbf y)
=\operatorname{Tr}[\rho_0E_{\gamma,\mathbf U}^{\boldsymbol\eta}(\mathbf y)]$.

Let $G_{\boldsymbol\eta,\mathbf U}(A_0,\ldots,A_{N-1})$ be the correlation
operator, which is independent of $\gamma$. A single-shot estimator
$f_{\boldsymbol\eta}(\mathbf y,\gamma)$, which does not depend on the unknown
initial state or free evolution, is averaged over independent repetitions. Universal
unbiasedness requires its mean to equal
$\operatorname{Tr}[\rho_0G_{\boldsymbol\eta,\mathbf U}]$ for every
$\rho_0,\mathbf U$, equivalently \cite{dubin2014,dressel2012}
\begin{equation}\label{eq:unbiased-original}
\int f_{\boldsymbol\eta}(\mathbf y,\gamma)
E_{\gamma,\mathbf U}^{\boldsymbol\eta}(\mathbf y)\,d\mathbf y
=G_{\boldsymbol\eta,\mathbf U}.
\end{equation}
The operators $G_{\boldsymbol\eta,\mathbf U}$ form a family indexed by
$\mathbf U$. Let $\mathcal{C}_\gamma^{\boldsymbol\eta}$ contain those families
for which one estimator satisfies Eq.~\eqref{eq:unbiased-original} for every $\mathbf U$.

For a state $\rho$ immediately before the $j$th measurement,
averaging over the measurement outcomes gives
\begin{equation}\label{eq:decoherence}
\int_{\mathbb R}
K_{j,\gamma}^{\eta_j}(y)\rho
K_{j,\gamma}^{\eta_j}(y)^\dagger\,dy
={}\sum_{a,b}e^{-\gamma^2(\lambda_{j,a}-\lambda_{j,b})^2/8}
P_{j,a}\rho P_{j,b}.
\end{equation}
This expression shows that dephasing induced by measurement backaction can bias correlation estimates. We introduce and record spectral phase control so that estimation biases associated with different control branches cancel upon averaging over the branches, thereby achieving unbiasedness. Specifically, after each nonfinal position readout, we independently sample $V_{j,1}^{(1)}=I$ or $V_{j,a}^{(2)}=e^{i\pi P_{j,a}}$, with probabilities $q_{j,1}^{(1)}$ and $q_{j,a}^{(2)}$ satisfying $q_{j,1}^{(1)}+\sum_a q_{j,a}^{(2)}=1$. No additional control is applied after momentum readouts.
For nonfinal position readouts, the labels
$(\alpha_j,\beta_j)$ take values in
$\{(1,1)\}\cup\{(2,a):1\le a\le D_j\}$.
For momentum readouts and the final readout, set
$(\alpha_j,\beta_j)=(1,1)$, $T_{j,1}^{(1),\eta_j}=I$, and $q_{j,1}^{(1)}=1$.
The operation after measurement $j$ is
\[
T_{j,\beta_j}^{(\alpha_j),\eta_j}
=V_{j,\beta_j}^{(\alpha_j)}\delta_{\eta_j,x}+I\delta_{\eta_j,p},
\]
with the full control sequence denoted by
$\mathbf T_{\boldsymbol\beta}^{(\boldsymbol\alpha),\boldsymbol\eta}
=\big(T_{j,\beta_j}^{(\alpha_j),\eta_j}\big)_{j=0}^{N-2}$.
The joint effect density of the readout vector and recorded control sequence is
\begin{equation}\label{eq:controlled-effect}
E_{\gamma,\mathbf U}^{\boldsymbol\eta,\mathrm{ctrl}}
(\mathbf y,\mathbf T_{\boldsymbol\beta}^{(\boldsymbol\alpha),\boldsymbol\eta})
=(B^{\mathrm{ctrl}})^\dagger B^{\mathrm{ctrl}},
\end{equation}
where
\[
B^{\mathrm{ctrl}}=\overleftarrow{\prod}_{j=0}^{N-1}
\Bigl(\sqrt{q_{j,\beta_j}^{(\alpha_j)}}\,
U_{j+1:j}T_{j,\beta_j}^{(\alpha_j),\eta_j}
K_{j,\gamma}^{\eta_j}(y_j)\Bigr).
\]
Choose an estimator
$f_{\mathbf T_{\boldsymbol\beta}^{(\boldsymbol\alpha),\boldsymbol\eta}}(\mathbf y,\gamma)$
for each full control sequence. Universal unbiasedness for every $\mathbf U$ requires
\begin{equation}\label{eq:unbiased-control}
\sum_{\boldsymbol\alpha,\boldsymbol\beta}\int
f_{\mathbf T_{\boldsymbol\beta}^{(\boldsymbol\alpha),\boldsymbol\eta}}(\mathbf y,\gamma)
E_{\gamma,\mathbf U}^{\boldsymbol\eta,\mathrm{ctrl}}
(\mathbf y,\mathbf T_{\boldsymbol\beta}^{(\boldsymbol\alpha),\boldsymbol\eta})\,d\mathbf y
=G_{\boldsymbol\eta,\mathbf U}.
\end{equation}
Let $\mathcal{C}_\gamma^{\boldsymbol\eta,\mathrm{ctrl}}$ be the set of operator
families obtained by varying the control probabilities
$q_{j,\beta_j}^{(\alpha_j)}$ and estimators
$f_{\mathbf T_{\boldsymbol\beta}^{(\boldsymbol\alpha),\boldsymbol\eta}}$
in Eq.~\eqref{eq:unbiased-control}.
At nonfinal position readouts, choosing
$q_{j,\beta_j}^{(\alpha_j)}=\delta_{\alpha_j,1}$ and
$f_{\mathbf T_{\boldsymbol\beta}^{(\boldsymbol\alpha),\boldsymbol\eta}}(\mathbf y,\gamma)
=f_{\boldsymbol\eta}(\mathbf y,\gamma)
\prod_{\substack{j=0\\\eta_j=x}}^{N-2}\delta_{\alpha_j,1}$
recovers the conventional protocol, giving $\mathcal{C}_\gamma^{\boldsymbol\eta}\subseteq
\mathcal{C}_\gamma^{\boldsymbol\eta,\mathrm{ctrl}}$.

Furthermore, the operator families obtainable with the controlled protocol
do not depend on finite nonzero measurement strength:
\begin{equation}\label{eq:inclusion}
\mathcal{C}^{\boldsymbol\eta,\mathrm{ctrl}}= \mathcal{C}_\gamma^{\boldsymbol\eta,\mathrm{ctrl}}
\end{equation}
This allows the measurement strength to be adjusted and optimized while
preserving universal unbiasedness. The proof is given in the Supplemental Material.

\textit{Nested correlations and unbiased estimation.---}
We now estimate nested anticommutator and commutator correlations.
Define $C_A^x(\cdot)\equiv\{A,\cdot\}/2$ and
$C_A^p(\cdot)\equiv-\mathrm{i}[A,\cdot]/2$, corresponding to the position and momentum readouts, respectively
\cite{mitchison2007,johansen2009,weak-readout-note}.
Applying these maps in reverse time order from the last Heisenberg
operator $A_{N-1}(t_{N-1})$ gives
\begin{equation}\label{eq:nested-target}
G_{\boldsymbol\eta,\mathbf U}
=C_{A_0(t_0)}^{\eta_0}\circ\cdots\circ
C_{A_{N-2}(t_{N-2})}^{\eta_{N-2}}\circ C_{A_{N-1}(t_{N-1})}^{x}(I).
\end{equation}
For a nonzero correlation operator, we fix $\eta_{N-1}=x$, since $C^p_{A_{N-1}}(I)=0$.
For the given forward time order, linear combinations of these nested
correlations span the corresponding correlation functions on the two-branch
Schwinger--Keldysh contour \cite{haehl2019,kamenev2009}.

To determine whether any universally unbiased single-shot
estimator exists at finite strength, we solve
Eq.~\eqref{eq:unbiased-original}. The result is the following spectral condition.

\textit{Theorem 1 (Unbiased estimation).---}
Suppose each $A_j$ has at least two distinct eigenvalues.
At any fixed finite strength, a universally unbiased single-shot
estimator exists if and only if the following condition holds at each
nonfinal position readout. Whenever
$\lambda_{j,a}+\lambda_{j,a'}=\lambda_{j,b}+\lambda_{j,b'}\ne0$,
the eigenvalues satisfy
\begin{equation}\label{eq:spectral-condition}
(\lambda_{j,a}-\lambda_{j,a'})^2
=(\lambda_{j,b}-\lambda_{j,b'})^2.
\end{equation}
Spectra with two distinct eigenvalues satisfy this condition automatically.
No such spectral restriction applies to probe momentum readout: for every $0<\gamma<\infty$ and every system operator $X$, there exists an estimator $h_{j,1}^{(1),p}(y,\gamma)$ satisfying
\begin{equation}\label{eq:momentum-unbiased}
\int h_{j,1}^{(1),p}(y,\gamma)K_{j,\gamma}^{p}(y)^\dagger
XK_{j,\gamma}^{p}(y),dy=C_{A_j}^{p}(X),
\end{equation}
as proved in Supplemental Material. Physically, commutators can also be accessed through linear response to weak perturbations, requiring only a final system measurement. Probe momentum readout corresponds to an outcome-dependent unitary perturbation, connecting it to this response picture in the weak-coupling limit.

\textit{No-go theorem (Equally spaced spectra).---}
Under the assumptions of Theorem~1, suppose $A_j$ is measured with
probe position readout at a nonfinal time and has three distinct eigenvalues
$\lambda_{j,a}<\lambda_{j,b}<\lambda_{j,c}$ with
$\lambda_{j,a}+\lambda_{j,c}=2\lambda_{j,b}\ne0$.
Then, for every fixed finite $\gamma>0$, no universally unbiased
single-shot estimator satisfies Eq.~\eqref{eq:unbiased-original}.
This spectral obstruction occurs for many common observables, including spin components in systems with spin quantum number greater than one and bosonic number operators.

To remove this spectral obstruction, we consider one local estimator
$h_{j,\beta_j}^{(\alpha_j),\eta_j}(y,\gamma)$ per readout.
For each full control sequence, we consider the factorized estimators
\begin{equation}\label{eq:control-estimator}
f_{\mathbf T_{\boldsymbol\beta}^{(\boldsymbol\alpha),\boldsymbol\eta}}(\mathbf y,\gamma)
=\prod_{j=0}^{N-1}h_{j,\beta_j}^{(\alpha_j),\eta_j}(y_j,\gamma).
\end{equation}
At each nonfinal measurement with probe position readout, choose the
control probabilities and local estimators to reproduce the anticommutator
map for every $X$:
\begin{equation}\label{eq:local-calibration}
\begin{aligned}
&\sum_{\alpha_j,\beta_j}q_{j,\beta_j}^{(\alpha_j)}
\int h_{j,\beta_j}^{(\alpha_j),x}(y,\gamma)
K_{j,\gamma}^{x}(y)^\dagger V_{j,\beta_j}^{(\alpha_j)\dagger}
 X\\
&\qquad\times V_{j,\beta_j}^{(\alpha_j)}K_{j,\gamma}^{x}(y)\,dy
=C_{A_j}^{x}(X).
\end{aligned}
\end{equation}

Define $I_{j,ab}[h]\equiv\int h(y,\gamma)\exp[-\gamma^2(y-(\lambda_{j,a}+\lambda_{j,b})/2)^2/2]\,dy$.
Keeping only the identity branch recovers the conventional protocol. In this case, Eqs.~\eqref{eq:kraus-x} and \eqref{eq:local-calibration} require, for $1\le a\le b\le D_j$,
\begin{equation}\label{eq:identity-branch-calibration}
\frac{\gamma}{\sqrt{2\pi}}I_{j,ab}[h_{j,1}^{(1),x}]
=\frac{\lambda_{j,a}+\lambda_{j,b}}{2}e^{\frac{\gamma^2}{8}(\lambda_{j,a}-\lambda_{j,b})^2}.
\end{equation}
For pairs with the same nonzero eigenvalue sum, the left-hand sides are identical, whereas the right-hand sides also depend on the squared eigenvalue gaps. Compatibility therefore requires Eq.~\eqref{eq:spectral-condition}.

Including the phase-flip branches changes this condition to
\begin{equation}\label{eq:phase-branch-calibration}
\begin{aligned}
\frac{\lambda_{j,a}+\lambda_{j,b}}{2}e^{\frac{\gamma^2}{8}(\lambda_{j,a}-\lambda_{j,b})^2}
=\frac{\gamma}{\sqrt{2\pi}}I_{j,ab}\Bigg[q_{j,1}^{(1)}h_{j,1}^{(1),x}
\\
{}+\sum_{c=1}^{D_j}q_{j,c}^{(2)}h_{j,c}^{(2),x}(1-2\delta_{ac})(1-2\delta_{bc})\Bigg].
\end{aligned}
\end{equation}
The signs introduced by the phase flips make the combination of branch readout functions depend on the spectral pair. This allows the left-hand sides to be adjusted independently even for pairs with equal eigenvalue sums, resolving the conflicting constraints of the conventional protocol. As proved in the Supplemental Material, these additional degrees of freedom suffice for Eq.~\eqref{eq:local-calibration} to admit solutions for arbitrary finite spectra and every $0<\gamma<\infty$.

\textit{Mean-square error and sampling cost.}---The sampling cost depends
on bias and variance. For the conventional protocol, we consider the estimator
\cite{mitchison2007,dressel2018,altuntas2025}
\begin{equation}\label{eq:sampling-product}
f_{\boldsymbol\eta}(\mathbf y,\gamma)
=\prod_{\substack{0\le j\le N-1\\\eta_j=x}}y_j
\prod_{\substack{0\le j\le N-2\\\eta_j=p}}h_{j,1}^{(1),p}(y_j,\gamma).
\end{equation}
Its mean approaches $\operatorname{Tr}[\rho_0G_{\boldsymbol\eta,\mathbf U}]$
as $\gamma\to0$. At finite strength, the bias is caused by the nonfinal position readouts.
For $R$ independent repetitions, the sample mean is
$
\hat f
=
\sum_{r=1}^{R}
f_{\boldsymbol{\eta}}(\boldsymbol{y}^{(r)},\gamma)/R.
$

For the given state and evolution, the mean-square error
(MSE) of the sample mean is
\begin{equation}
\label{eq:sampling-mse-standard}
M_R\equiv\mathrm E[(\hat f-\operatorname{Tr}[\rho_0G_{\boldsymbol\eta,\mathbf U}])^2]
=\epsilon^2+\frac{\operatorname{Var}f_{\boldsymbol\eta}}R,
\end{equation}
where
$\epsilon=
\mathrm{E}
[f_{\boldsymbol{\eta}}]
-\operatorname{Tr}[\rho_0G_{\boldsymbol\eta,\mathbf U}]$
is the bias. We write $M_R(\hat f,\gamma;\rho_0,\mathbf U)$ when its
arguments are needed. At fixed $R$, $\rho_0$, and $\mathbf U$, we minimize
the MSE over the measurement strength of the conventional protocol:
\begin{equation}\label{eq:sampling-mse-optimal}
\mathcal M_R
\equiv\min_{\gamma'\in(0,\infty]}M_R(\hat f,\gamma';\rho_0,\mathbf U).
\end{equation}
The optimization includes the strong-measurement limit. We denote an optimal strength by
\[
\gamma'_*=\operatorname*{arg\,min}_{\gamma'\in(0,\infty]}
M_R(\hat f,\gamma';\rho_0,\mathbf U).
\]
When the bias is not identically zero, its expansion about $\gamma'=0$
starts at an even order $n\ge2$, determined by the state and evolution.
The weak strength limit then gives
$\epsilon^2\sim\gamma'^{2n}$ and
$\operatorname{Var}f_{\boldsymbol\eta}\sim\gamma'^{-2N}$.
If the global optimum approaches this limit as $R\to\infty$, balancing
variance and squared bias gives
$\gamma'_*\sim R^{-1/[2(N+n)]}$ and
\begin{equation}\label{eq:sampling-optimal-scaling}
\mathcal M_R\sim R^{-n/(N+n)}.
\end{equation}
The derivation of this asymptotic scaling is given in the Supplemental Material.

This asymptotic result need not apply if the bias accidentally vanishes outside the weak strength regime. The collective-spin example below exhibits
these situations.

The controlled protocol always admits a universally unbiased estimator.
Its sample-mean MSE is
\begin{equation}\label{eq:sampling-mse-controlled}
M_R^{\mathrm{ctrl}}
=\frac{\operatorname{Var}f_{\mathbf T_{\boldsymbol\beta}^{(\boldsymbol\alpha),\boldsymbol\eta}}}{R}.
\end{equation}
Write $\mathbf q=\{q_{j,\beta}^{(\alpha)}\}$ and
$h=\{h_{j,\beta}^{(\alpha),\eta_j}\}_{j=0}^{N-1}$, with $\mathbf q_j,h_j$
denoting the collections at time $j$.
We select the probabilities $\mathbf q$, local estimators $h$, and strength
$\gamma$ before specifying the state or evolution:
\begin{equation}\label{eq:joint-minimax}
(\mathbf q_*,h_*,\gamma_*)=\operatorname*{arg\,min}_{\mathbf q,h,\gamma\in(0,\infty]}
\max_{\rho_0,\mathbf U}M_R^{\mathrm{ctrl}}.
\end{equation}
The minimization is subject to Eqs.~\eqref{eq:momentum-unbiased} and
\eqref{eq:local-calibration}.
The corresponding actual MSE
$\mathcal M_R^{\mathrm{ctrl}}
=\left.M_R^{\mathrm{ctrl}}\right|_{(\mathbf q,h,\gamma)=(\mathbf q_*,h_*,\gamma_*)}$
scales as $R^{-1}$ because the estimator is unbiased.

We finally compare $\mathcal M_R^{\mathrm{ctrl}}$ with $\mathcal M_R$ for the
conventional protocol at the same initial state and evolution. A crossover
sample size, when present, satisfies
\begin{equation}\label{eq:sampling-crossover}
\mathcal M_{R_\times}=\mathcal M_{R_\times}^{\mathrm{ctrl}}.
\end{equation}
Below the crossover the conventional protocol has the smaller MSE;
above it the controlled protocol does. When the conventional optimum
has the weak-strength asymptotics in Eq.~\eqref{eq:sampling-optimal-scaling},
the controlled protocol is favored at sufficiently large $R$.

\textit{Collective-spin temporal correlations.---}
We consider a generalized Lipkin--Meshkov--Glick (LMG)
model~\cite{lipkin1965} with Hamiltonian
\begin{equation}\label{eq:spin-model}
H_J=\Omega J_x+\Delta J_z+\frac{\kappa}{J}J_z^2.
\end{equation}
This Hamiltonian also describes two-mode bosonic dynamics~\cite{raghavan1999,japha2016}.
We take the initial state to be a spin coherent state
$\rho_0=|\psi_J\rangle\langle\psi_J|$, where
$|\psi_J\rangle=e^{-i\varphi J_z}e^{-i\theta J_y}|J,J\rangle$.
We measure $A_0=A_1=A=J_z/J$ at times $0$ and $t$ and set
$\tau=\Omega t$. For probe-position readouts at both times,
$\boldsymbol\eta=(x,x)$, and the target is
$G_{xx}=\{A,A(t)\}/2$, with $A(t)=e^{iH_Jt}Ae^{-iH_Jt}$.
For the conventional protocol, we use $f_{xx}=y_0y_1$,
with bias
$\epsilon=\mathrm E[f_{xx}]-\operatorname{Tr}(\rho_0G_{xx})$.
For a finite bias exponent $n$, its squared bias has the following scaling as $\gamma'\to0$:
\begin{equation}\label{eq:examplebias}
\epsilon^2=\left[\sum_{k=1}^{2J-1}w_k
\left(e^{-\gamma'^2k^2/(8J^2)}-1\right)\right]^2
\propto\frac{\gamma'^{2n}}{J^{2n}}
\left(\sum_{k=1}^{2J-1}k^nw_k\right)^2,
\end{equation}
where $w_k=\sum_{|m-m'|=k}\rho_{0,m'm}
(m+m')A_{mm'}(t)/(2J)$.
The first nonzero even moment $\sum_k k^nw_k$ determines $n$.
The $2J-1$ distinct gaps therefore allow $n=2,4,\ldots,4J-2$;
$n=\infty$ requires every $w_k=0$. These are spectral possibilities:
the dynamics in Eq.~\eqref{eq:spin-model} can restrict which ones occur.

\begin{figure}[t]
\centering
\includegraphics[width=\columnwidth]{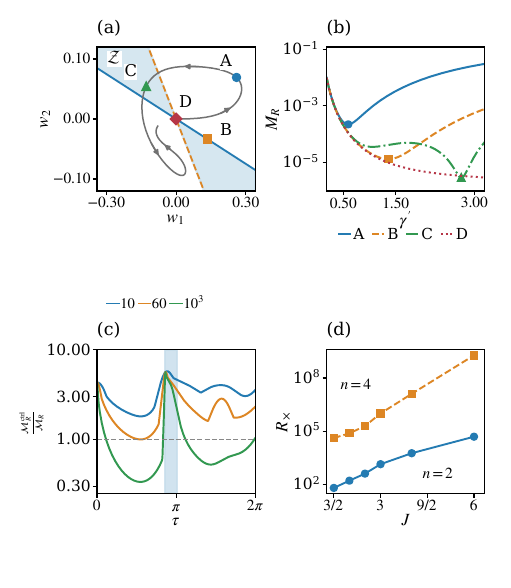}
\caption{Collective-spin results for $\Delta/\Omega=-1$,
$\kappa/\Omega=3/2$, and $\theta=\pi/3$.
(a) $J=3/2$ bias diagram in $(w_1,w_2)$; $\mathcal Z$ admits finite-strength unbiasedness.
The gray curve shows the $\varphi=\pi/4$ trajectory in (c),
$0\le\tau\le2\pi$; arrows indicate increasing $\tau$.
A $(\tau,\varphi)=(\pi/2,\pi/4)$ and C $(3,\pi/4)$ have $n=2$.
B $(\pi/2,\varphi_c)$, with $\varphi_c\simeq1.85$, has $n=4$.
D at $(\tau,\varphi)\simeq(2.38,1.51)$ has $n=\infty$.
(b) MSE for the conventional protocol at $R=10^5$.
Symbols mark MSE minima attained at finite measurement strength.
(c) Ratio $\mathcal M_R^{\mathrm{ctrl}}/\mathcal M_R$ for $R=10,60,10^3$
at $\varphi=\pi/4$.
Shading marks the corresponding time slice of $\mathcal Z$ in (a).
(d) Crossover sample size versus $J$ at $\tau=\pi/2$:
$n=2$ at $\varphi=\pi/4$ and $n=4$ at the phase satisfying
$\sum_k k^2w_k=0$.}
\label{fig:spin-bias-cost}
\end{figure}

For $A=J_z/J$, the smallest spin quantum number violates the spectral condition in Theorem~1 is $J=3/2$. In this case,
$\epsilon=(e^{-\gamma'^2/18}-1)w_1+(e^{-2\gamma'^2/9}-1)w_2$. Figure~\ref{fig:spin-bias-cost}(a)
shows the region $\mathcal Z$ where the estimator used in the conventional
protocol is unbiased at a finite strength:
$w_2\ne0$ and $1<-w_1/w_2<4$.
The optimal measurement strength approaches the weak and strong limits at the boundaries $w_1+4w_2=0$ and $w_1+w_2=0$, respectively.
Point B has $n=4$, and their intersection D has $n=\infty$.
Points A and C have $n=2$, with only C inside $\mathcal Z$.

For the controlled protocol, we determine $\mathbf q_*,h_*,\gamma_*$
from Eq.~\eqref{eq:joint-minimax}. In particular, when the spectrum
of $A_{N-1}$ is symmetric about zero, as for $A=J_z/J$,
the optimization reduces to local problems:
\begin{equation}\label{eq:local-mse-bound}
\min_{\mathbf q,h,\gamma}\max_{\rho_0,\mathbf U}
\operatorname{Var}f_{\mathbf T_{\boldsymbol\beta}^{(\boldsymbol\alpha),\boldsymbol\eta}}
=\min_{\gamma}\prod_{j=0}^{N-1}\sigma_{j,*}^{\eta_j}(\gamma),
\end{equation}
where $\sigma_{j,*}^{\eta_j}(\gamma)\equiv
\min_{\mathbf q_j,h_j}\max_a\mathrm E_{\lambda_{j,a}}
[(h_{j,\beta}^{(\alpha),\eta_j}(y,\gamma))^2]$.
The expectation includes the local readout and sampled control for an
input eigenstate of eigenvalue $\lambda_{j,a}$.
Details of the local optimization are given in the Supplemental Material.

For $J=3/2$, we use the ordered spectrum $(-1,-1/3,1/3,1)$ and solve
Eq.~\eqref{eq:local-mse-bound} numerically, including both local estimators,
to fix one protocol for all initial states and evolutions. All controlled-protocol results for $J=3/2$ use this same set of probabilities, estimators, and measurement strength.

Figure~\ref{fig:spin-bias-cost}(b) compares the MSEs of the estimator
used in the conventional protocol at A--D.
The MSE minimum is at finite strength for A--C and in the
strong-measurement limit for D. The accidental cancellations at C and D
illustrate the situations following Eq.~\eqref{eq:sampling-optimal-scaling}.
Figure~\ref{fig:spin-bias-cost}(c) compares the two optimized MSEs for
different sample sizes. Unbiasedness achieved by a finite measurement strength at C favors the
conventional protocol nearby, and increasing $R$
expands the time intervals favoring the controlled protocol.
The shaded interval is the $\varphi=\pi/4$ time slice of $\mathcal Z$.
For the controlled protocol, $\mathcal M_R^{\mathrm{ctrl}}\sim R^{-1}$.
Figure~\ref{fig:spin-bias-cost}(d) shows the crossover in
Eq.~\eqref{eq:sampling-crossover} for $n=2$ and $n=4$, with
$\mathcal M_R\sim R^{-n/(n+2)}$ in the weak strength regime.
For each $J$, the probabilities $\mathbf q_*$, local estimators $h_*$,
and strength $\gamma_*$ are fixed by Eq.~\eqref{eq:joint-minimax}.
Smaller systems have lower crossover sample sizes in these examples, since
fewer distinct eigenvalues require fewer spectral-flip branches, reducing
the additional fluctuations introduced by random control.

\begin{figure}[t]
\centering
\includegraphics[width=\columnwidth]{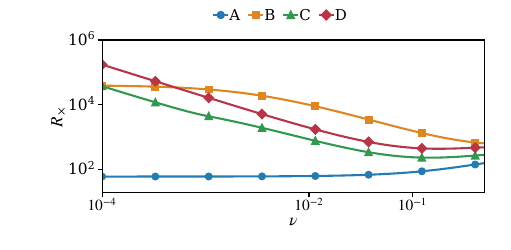}
\caption{Crossover sample size under independent quenched fluctuations
of $\Omega$ and $\Delta$ for $J=3/2$, with $\kappa$ and physical $t$
fixed. A--D are the nominal working points of
Fig.~\ref{fig:spin-bias-cost}. The control probabilities, local estimators, and strength from Eq.~\eqref{eq:joint-minimax} are fixed for
all curves and disorder variances $\nu$.}
\label{fig:spin-windows}
\end{figure}

To test whether bias cancellation survives uncertainty in the system
dynamics, we introduce quenched fluctuations in the two fields:
$\Omega\to\Omega(1+\xi_\Omega)$ and
$\Delta\to\Delta(1+\xi_\Delta)$.
The variables $\xi_\Omega$ and $\xi_\Delta$ are independent Gaussians
with zero mean and variance $\nu$, fixed over the $R$ measurement
sequences of each realization. The interaction $\kappa$ and physical evolution time $t$
are held fixed. For $\xi=(\xi_\Omega,\xi_\Delta)$, the actual evolution
is $U_\xi=e^{-itH_J^{(\xi)}}$. The estimated quantity is the
realization's actual correlation
$C_\xi=\operatorname{Tr}[\rho_0\{A,U_\xi^\dagger AU_\xi\}]/2$.
Let an overbar denote the average over quenched realizations. For $J=3/2$,
the quenched squared bias is
$\overline{\epsilon_\xi^2}
=\overline{\bigl[(e^{-\gamma'^2/18}-1)w_1(\xi)+(e^{-2\gamma'^2/9}-1)w_2(\xi)\bigr]^2}$.
The conditional bias is
$\epsilon_\xi=\mathrm E[f_{xx}\mid\xi]-C_\xi$.
Field fluctuations destroy even D's accidental bias cancellation.
The local unbiasedness conditions hold for every $U_\xi$,
so the estimator used in the controlled protocol remains unbiased in each realization.
Figure~\ref{fig:spin-windows} shows the crossovers for A--D, defined by
$\overline{\mathcal M_{R_\times}^{\mathrm{ctrl}}}
=\min_{\gamma'}\overline{M_{R_\times}}$.

\textit{Discussion.---}
The spectra of the measured observables govern the limits imposed by measurement backaction on universal unbiased estimation of temporal correlations. We show that random spectral phase
control can remove the resulting bias at arbitrary finite measurement strength.
The protocol applies to general finite-spectrum Hermitian observables
without knowledge of the initial state or system dynamics.
The control probabilities, local estimators, and measurement strength can be optimized in advance, and the random control sequences can then be generated without real-time feedback. Our collective-spin example shows how spectrum-only optimization reduces
the sampling cost and how quenched Hamiltonian disorder can favor the
controlled measurement protocol.

Existing experimental techniques may provide routes to the required
spectrally selective phase flips.  For example, selective number-dependent arbitrary phase (SNAP) gates in superconducting cavities~\cite{heeres2015} and their virtual counterparts for a high-spin nucleus in silicon~\cite{yu2025} allow independent phase control of different basis states, providing a possible route to selective flips of arbitrary spectral eigenspaces.

Since sequential measurements transfer information from the system to the probes, a natural extension is to optimize feedback control conditioned on earlier measurement outcomes. Another direction is to develop joint estimators using multiple measurement records.

\textit{Acknowledgment.}---This work was supported by the
Science Challenge Project (Grant No. TZ2025017).


\begin{thebibliography}{40}
\bibitem{weak-readout-note}
For the Gaussian probes in Eq.~\eqref{eq:kraus}, one possible choice
of local estimators in the conventional protocol is $y$ for probe position
readouts and $-2y/\gamma$ for probe momentum readouts. In the
weak-measurement limit, these estimators give
\[
\begin{aligned}
\lim_{\gamma\to0}\int_{\mathbb R} y\,&
K_{j,\gamma}^{x}(y)^\dagger\cdot K_{j,\gamma}^{x}(y)\,dy=\frac{\{A_j,\cdot\}}{2},
\\
\lim_{\gamma\to0}\int_{\mathbb R}\frac{-2y}{\gamma}\,&
K_{j,\gamma}^{p}(y)^\dagger\cdot K_{j,\gamma}^{p}(y)\,dy=-\frac{i}{2}[A_j,\cdot].
\end{aligned}
\]
At finite strength, the estimators for probe momentum readouts in
Eq.~\eqref{eq:sampling-product} reproduce the commutator map exactly.
\end{thebibliography}
\end{document}


\title{Supplemental Material for\texorpdfstring{\\}{ }
Overcoming Backaction-Induced Bias in Quantum Temporal Correlations}
\author{Yi-Mu Du}
\affiliation{Graduate School of China Academy of Engineering Physics, Beijing 100193, China}
\maketitle
\relpenalty=70\binoppenalty=150
\tolerance=3000\emergencystretch=2em

This Supplemental Material is organized as follows. Section~\ref{sec:space} characterizes the operators estimable with the controlled protocol and proves that their space is independent of finite measurement strength. Section~\ref{sec:theorem} proves Theorem~1. Section~\ref{sec:expansions} derives the bias expansion of factorized estimators and their variance in the weak-measurement limit. Section~\ref{sec:optimization} optimizes the control probabilities, local estimators, and common measurement strength.

\section{Operators estimable with control}
\label{sec:space}

Following the notation of the main text, let $\mathcal C_\gamma^{\boldsymbol\eta,\mathrm{ctrl}}$ denote the space of operators $G_{\boldsymbol\eta,\mathbf U}$ estimable with the controlled protocol at strength $\gamma$. Here a single estimator must have mean $\Tr(\rho_0 G_{\boldsymbol\eta,\mathbf U})$ for every initial state $\rho_0$ and every free evolution $\mathbf U$. We show that all finite strengths $0<\gamma,\gamma'<\infty$ yield the same space, denoted by $\mathcal C^{\boldsymbol\eta,\mathrm{ctrl}}$:
\begin{equation}\label{eq:strength}
\mathcal C_\gamma^{\boldsymbol\eta,\mathrm{ctrl}}=\mathcal C^{\boldsymbol\eta,\mathrm{ctrl}}=\mathcal C_{\gamma'}^{\boldsymbol\eta,\mathrm{ctrl}}.
\end{equation}

\subsection{Finite sums of factorized estimators}\label{subsec:factorized-reduction}

Fix $0<\gamma<\infty$ and the readout type $\eta_j$ (position or momentum) at each measurement, with $\boldsymbol\eta=(\eta_0,\ldots,\eta_{N-1})$. A complete control sequence is $\mathbf T=\mathbf T_{\boldsymbol\beta}^{(\boldsymbol\alpha),\boldsymbol\eta}$, with probability $\prod_{j=0}^{N-1}q_{j,\beta_j}^{(\alpha_j)}$. As in the main text, $q_{j,1}^{(1)}=1$ at momentum readouts and the final readout.

We consider complex-valued estimators satisfying $\mathrm E_{\rho_0,\mathbf U}|f_{\mathbf T}(\mathbf y,\gamma)|<\infty$ for every initial state $\rho_0$ and free evolution $\mathbf U$. For each $\mathbf U$, the corresponding operator is
\begin{equation}\label{eq:joint-estimator-family}
G_{\boldsymbol\eta,\mathbf U}=\sum_{\boldsymbol\alpha,\boldsymbol\beta}\int f_{\mathbf T}(\mathbf y,\gamma)E_{\gamma,\mathbf U}^{\boldsymbol\eta,\mathrm{ctrl}}(\mathbf y,\mathbf T)\,d\mathbf y.
\end{equation}

For a factorized estimator, the dependence on the control sequence $\mathbf T$ and the readout vector $\mathbf y$ takes the product form
\begin{equation}\label{eq:factorized-estimator}
f_{\mathbf T}(\mathbf y,\gamma)=\prod_{j=0}^{N-1}h_{j,\beta_j}^{(\alpha_j),\eta_j}(y_j,\gamma).
\end{equation}

For each estimator in Eq.~\eqref{eq:joint-estimator-family}, we now construct a finite sum of factorized estimators in Eq.~\eqref{eq:factorized-estimator}. The sum gives the same $G_{\boldsymbol\eta,\mathbf U}$ for every $\mathbf U$,
\[
\sum_{\boldsymbol\alpha,\boldsymbol\beta}\int
\left[\sum_r f_{\mathbf T}^{(r)}(\mathbf y,\gamma)\right]
E_{\gamma,\mathbf U}^{\boldsymbol\eta,\mathrm{ctrl}}(\mathbf y,\mathbf T)\,d\mathbf y
=G_{\boldsymbol\eta,\mathbf U}.
\]
To prove this, we examine how $E_{\gamma,\mathbf U}^{\boldsymbol\eta,\mathrm{ctrl}}$ depends on $\mathbf y$. Recall $A_j=\sum_{a=1}^{D_j}\lambda_{j,a}P_{j,a}$. To fix notation, let $\chi_{j,ab}^x=(\lambda_{j,a}+\lambda_{j,b})/2$ and $\chi_{j,ab}^p=-i(\lambda_{j,a}-\lambda_{j,b})/2$.

For the position measurement, write $\phi_{\vartheta,\gamma}(y)=\gamma e^{-\gamma^2(y-\vartheta)^2/2}/\sqrt{2\pi}$ for the Gaussian density centered at $\vartheta$, and define $1+r_{j,ab}=e^{\gamma^2(\lambda_{j,a}-\lambda_{j,b})^2/8}$. Substituting the spectral decomposition of $K_{j,\gamma}^x$ gives
\begin{equation}\label{eq:instrument-expansion}
\begin{aligned}
&K_{j,\gamma}^x(y)^\dagger XK_{j,\gamma}^x(y)=\sum_{a,b}k_{j,\gamma}^x(y;a,b)P_{j,a}XP_{j,b},\\
&k_{j,\gamma}^x(y;a,b)=\frac{\phi_{\chi_{j,ab}^x,\gamma}(y)}{1+r_{j,ab}}.
\end{aligned}
\end{equation}

The corresponding expansion for the momentum measurement is
\begin{equation}\label{eq:kernel-p}
\begin{aligned}
&K_{j,\gamma}^p(y)^\dagger XK_{j,\gamma}^p(y)=\sum_{a,b}k_{j,\gamma}^p(y;a,b)P_{j,a}XP_{j,b},\\
&k_{j,\gamma}^p(y;a,b)=\sqrt{2/\pi}e^{-2y^2}e^{-2\gamma\chi_{j,ab}^p y}.
\end{aligned}
\end{equation}

We now compose the maps in measurement order. For a sequence of spectral indices $\mathbf a=(a_0,\ldots,a_{N-1})$ and a fixed control sequence $\mathbf T$, define
\begin{equation}\label{eq:spectral-path}
Q_{\mathbf a,\mathbf U}^{\mathbf T}:=\overleftarrow{\prod}_{j=0}^{N-1}\left[U_{j+1:j}T_{j,\beta_j}^{(\alpha_j),\eta_j}P_{j,a_j}\right].
\end{equation}
The operators $Q_{\mathbf a,\mathbf U}^{\mathbf T}$ contain all control and evolution factors and are independent of $\mathbf y$. The joint effect density for the recorded values $\mathbf y$ and the control sequence $\mathbf T$ therefore has the expansion
\begin{equation}\label{eq:spectral-effect}
E_{\gamma,\mathbf U}^{\boldsymbol\eta,\mathrm{ctrl}}(\mathbf y,\mathbf T)
=\left(\prod_{j=0}^{N-1}q_{j,\beta_j}^{(\alpha_j)}\right)
\sum_{\mathbf a,\mathbf b}\left[\prod_{j=0}^{N-1}k_{j,\gamma}^{\eta_j}(y_j;a_j,b_j)\right]
\left(Q_{\mathbf a,\mathbf U}^{\mathbf T}\right)^\dagger Q_{\mathbf b,\mathbf U}^{\mathbf T}.
\end{equation}
Equation~\eqref{eq:spectral-effect} expresses the $\mathbf y$ dependence of $E_{\gamma,\mathbf U}^{\boldsymbol\eta,\mathrm{ctrl}}$ through products of $k_{j,\gamma}^{\eta_j}(y_j;a_j,b_j)$.

To replace a general estimator by a sum of products, we construct an orthonormal basis for these functions, using the index sets
\begin{equation}\label{eq:spectral-sets}
\begin{aligned}
&\mathcal S_j^x=\{\chi_{j,ab}^x:1\le a\le b\le D_j\},\\
&\mathcal S_j^p=\{\lambda_{j,a}-\lambda_{j,b}:1\le a,b\le D_j\}.
\end{aligned}
\end{equation}
For position readouts, use the functions $\phi_{\vartheta,\gamma}(y)$ with $\vartheta\in\mathcal S_j^x$. For momentum readouts, use $\sqrt{2/\pi}e^{-2y^2}e^{i\gamma\omega y}$ with $\omega\in\mathcal S_j^p$. For $\vartheta,\vartheta'\in\mathcal S_j^x$, the position Gram matrix is
\begin{equation}\label{eq:Gx}
\mathsf G_{j,\gamma}^x(\vartheta,\vartheta')=\int\phi_{\vartheta,\gamma}(y)\phi_{\vartheta',\gamma}(y)\,dy=\frac{\gamma}{2\sqrt\pi}e^{-\gamma^2(\vartheta-\vartheta')^2/4}.
\end{equation}
For $\omega,\omega'\in\mathcal S_j^p$, the momentum Gram matrix is
\begin{equation}\label{eq:Gp}
\mathsf G_{j,\gamma}^p(\omega,\omega')=\frac2\pi\int e^{-4y^2}e^{i\gamma(\omega'-\omega)y}\,dy=\frac1{\sqrt\pi}e^{-\gamma^2(\omega-\omega')^2/16}.
\end{equation}
The functions $\phi_{\vartheta,\gamma}$ with distinct $\vartheta\in\mathcal S_j^x$ are linearly independent, as are $e^{i\gamma\omega y}$ with distinct $\omega\in\mathcal S_j^p$. Both Gram matrices are therefore positive definite.

For $\vartheta\in\mathcal S_j^\eta$ and $\eta=x,p$, multiplying the corresponding functions by the inverse square roots of these matrices defines
\begin{equation}\label{eq:orthonormal-readout}
\begin{aligned}
&\mathsf u_{j,\vartheta}^x(y;\gamma)=\sum_{\vartheta'\in\mathcal S_j^x}\phi_{\vartheta',\gamma}(y)(\mathsf G_{j,\gamma}^x)^{-1/2}(\vartheta',\vartheta),\\
&\mathsf u_{j,\vartheta}^p(y;\gamma)=\sqrt{2/\pi}e^{-2y^2}\sum_{\omega\in\mathcal S_j^p}e^{i\gamma\omega y}(\mathsf G_{j,\gamma}^p)^{-1/2}(\omega,\vartheta).
\end{aligned}
\end{equation}
These functions are orthonormal: $\int[\mathsf u_{j,\vartheta}^{\eta_j}(y;\gamma)]^*\mathsf u_{j,\vartheta'}^{\eta_j}(y;\gamma)\,dy=\delta_{\vartheta,\vartheta'}$. Each $k_{j,\gamma}^{\eta_j}(\cdot;a,b)$ lies in their span, so projecting it onto that span leaves it unchanged:
\begin{equation}\label{eq:dual-local}
\int\left[\sum_{\vartheta\in\mathcal S_j^{\eta_j}}\mathsf u_{j,\vartheta}^{\eta_j}(y;\gamma)[\mathsf u_{j,\vartheta}^{\eta_j}(y';\gamma)]^*\right]k_{j,\gamma}^{\eta_j}(y';a,b)\,dy'=k_{j,\gamma}^{\eta_j}(y;a,b).
\end{equation}

For the joint readout, let $\boldsymbol\vartheta=(\vartheta_0,\ldots,\vartheta_{N-1})\in\prod_{j=0}^{N-1}\mathcal S_j^{\eta_j}$. Applying Eq.~\eqref{eq:dual-local} to each $k_{j,\gamma}^{\eta_j}$ in Eq.~\eqref{eq:spectral-effect}, while keeping the control sequence $\mathbf T$ fixed, gives
\begin{equation}\label{eq:dual-joint}
E_{\gamma,\mathbf U}^{\boldsymbol\eta,\mathrm{ctrl}}(\mathbf y,\mathbf T)
=\sum_{\boldsymbol\vartheta}\left[\prod_{j=0}^{N-1}\mathsf u_{j,\vartheta_j}^{\eta_j}(y_j;\gamma)\right]
\int\left[\prod_{j=0}^{N-1}[\mathsf u_{j,\vartheta_j}^{\eta_j}(y'_j;\gamma)]^*\right]
E_{\gamma,\mathbf U}^{\boldsymbol\eta,\mathrm{ctrl}}(\mathbf y',\mathbf T)\,d\mathbf y'.
\end{equation}

For a control sequence of nonzero probability, define $\widetilde f_{\mathbf T}$ by
\begin{equation}\label{eq:factorized-replacement}
\widetilde f_{\mathbf T}(\mathbf y,\gamma)=\sum_{\boldsymbol\vartheta}\left[\int f_{\mathbf T}(\mathbf y',\gamma)\prod_{j=0}^{N-1}\mathsf u_{j,\vartheta_j}^{\eta_j}(y'_j;\gamma)\,d\mathbf y'\right]\prod_{j=0}^{N-1}[\mathsf u_{j,\vartheta_j}^{\eta_j}(y_j;\gamma)]^*.
\end{equation}
For a control sequence of zero probability, set $\widetilde f_{\mathbf T}=0$ because its effect density vanishes. Substituting Eq.~\eqref{eq:dual-joint} into the integral with $f_{\mathbf T}$, and using the definition of $\widetilde f_{\mathbf T}$ in Eq.~\eqref{eq:factorized-replacement}, gives, for every $\mathbf U$,
\begin{equation}\label{eq:same-operator-family}
\int\widetilde f_{\mathbf T}(\mathbf y,\gamma)E_{\gamma,\mathbf U}^{\boldsymbol\eta,\mathrm{ctrl}}(\mathbf y,\mathbf T)\,d\mathbf y
=\int f_{\mathbf T}(\mathbf y,\gamma)E_{\gamma,\mathbf U}^{\boldsymbol\eta,\mathrm{ctrl}}(\mathbf y,\mathbf T)\,d\mathbf y.
\end{equation}
Summing over the control sequences shows that $\widetilde f$ and $f$ give the same $G_{\boldsymbol\eta,\mathbf U}$ for every $\mathbf U$. By Eq.~\eqref{eq:factorized-replacement}, $\widetilde f$ is a finite sum of factorized estimators of the form in Eq.~\eqref{eq:factorized-estimator}.

It remains to determine which maps these local factors can generate. For control branch $(\alpha_j,\beta_j)$ and readout $y$ at time $j$, the corresponding map is
\begin{equation}\label{eq:record-map}
\mathcal K_{j,\gamma,\beta_j}^{(\alpha_j),\eta_j}(y)(X)=K_{j,\gamma}^{\eta_j}(y)^\dagger\big(T_{j,\beta_j}^{(\alpha_j),\eta_j}\big)^\dagger XT_{j,\beta_j}^{(\alpha_j),\eta_j}K_{j,\gamma}^{\eta_j}(y).
\end{equation}
With $\mathcal U_{j+1:j}(X)=U_{j+1:j}^\dagger XU_{j+1:j}$, composing these maps in measurement order gives the joint effect density
\begin{equation}\label{eq:record-effect}
E_{\gamma,\mathbf U}^{\boldsymbol\eta,\mathrm{ctrl}}(\mathbf y,\mathbf T)=\left(\prod_{j=0}^{N-1}q_{j,\beta_j}^{(\alpha_j)}\right)\left(\mathop{\bigcirc}_{j=0}^{N-1}\left[\mathcal K_{j,\gamma,\beta_j}^{(\alpha_j),\eta_j}(y_j)\circ\mathcal U_{j+1:j}\right]\right)(I).
\end{equation}
The maps in Eq.~\eqref{eq:record-effect} are ordered by increasing $j$ from left to right and act on $I$ from right to left.

At time $j$, integrate the readout map against the local factor for each control branch and sum over branches with their probabilities. For any operator $X$, the resulting map has the spectral expansion
\begin{equation}\label{eq:local-space-definition}
\sum_{\alpha_j,\beta_j}q_{j,\beta_j}^{(\alpha_j)}
\int h_{j,\beta_j}^{(\alpha_j),\eta_j}(y,\gamma)
\mathcal K_{j,\gamma,\beta_j}^{(\alpha_j),\eta_j}(y)(X)\,dy
=\sum_{a,b}\mathsf s_{j,ab}P_{j,a}XP_{j,b}.
\end{equation}
Varying the allowed control probabilities and the bounded complex-valued local factors gives a space of maps, denoted by $\mathscr L_{j,\gamma}^{\eta_j,\mathrm{ctrl}}$. Let $\mathcal P_j^{\eta_j}$ denote the corresponding space of coefficient matrices $\mathsf s_j=[\mathsf s_{j,ab}]_{a,b=1}^{D_j}$. We now determine the coefficient spaces for position and momentum readouts and show that they are independent of $\gamma$ for $0<\gamma<\infty$.

\subsection{Position readout}\label{subsec:position-space}

For position readouts, we show that the coefficient space is the space of complex symmetric matrices,
\[
\mathcal P_j^x=\{\mathsf s_j\in\mathbb C^{D_j\times D_j}:\mathsf s_{j,ab}=\mathsf s_{j,ba}\}.
\]

First fix positive probabilities satisfying $q_{j,1}^{(1)}+\sum_{\ell=1}^{D_j}q_{j,\ell}^{(2)}=1$, and write $(hq)_{j,\beta}^{(\alpha)}=h_{j,\beta}^{(\alpha),x}q_{j,\beta}^{(\alpha)}$. Expanding $\mathcal L_{j,\gamma}^x[hq](X)$ in $P_{j,a}XP_{j,b}$ gives
\begin{equation}\label{eq:local-controlled}
\begin{aligned}
&\mathcal L_{j,\gamma}^x[hq](X)=\int(hq)_{j,1}^{(1)}K_{j,\gamma}^x(y)^\dagger XK_{j,\gamma}^x(y)\,dy\\
&+\sum_{\ell=1}^{D_j}\int(hq)_{j,\ell}^{(2)}K_{j,\gamma}^x(y)^\dagger(V_{j,\ell}^{(2)})^\dagger XV_{j,\ell}^{(2)}K_{j,\gamma}^x(y)\,dy=\sum_{a,b}\mathsf s_{j,ab}P_{j,a}XP_{j,b},\\
&\mathsf s_{j,ab}=\frac1{1+r_{j,ab}}\int\phi_{\chi_{j,ab}^x,\gamma}(y)\left[(hq)_{j,1}^{(1)}(y,\gamma)+\sum_{\ell=1}^{D_j}(hq)_{j,\ell}^{(2)}(y,\gamma)(1-2\delta_{a\ell})(1-2\delta_{b\ell})\right]dy.
\end{aligned}
\end{equation}
The expression for $\mathsf s_{j,ab}$ is invariant under $a\leftrightarrow b$, so every realizable coefficient matrix is symmetric.

To construct the local factors for a prescribed symmetric matrix, we use the inverse of the Gram matrix defined in Eq.~\eqref{eq:Gx}. For $\vartheta\in\mathcal S_j^x$, define
\begin{equation}\label{eq:pair}
\widetilde\phi_{j,\vartheta}^x(y,\gamma)=\sum_{\vartheta'\in\mathcal S_j^x}(\mathsf G_{j,\gamma}^x)^{-1}(\vartheta',\vartheta)\phi_{\vartheta',\gamma}(y).
\end{equation}
These functions are real and bounded at fixed $\gamma$, and satisfy $\int\widetilde\phi_{j,\vartheta}^x(y,\gamma)\phi_{\vartheta',\gamma}(y)\,dy=\delta_{\vartheta,\vartheta'}$.

For any prescribed symmetric matrix $\mathsf s_j$, write the product of each branch probability and its local factor as
\[
(hq)_{j,\beta}^{(\alpha)}(y,\gamma)=\sum_{1\le c\le d\le D_j}(hq)_{j,\beta}^{(\alpha;cd)}(y,\gamma)
\]
with the following terms chosen to satisfy Eq.~\eqref{eq:local-controlled} for $1\le c\le d\le D_j$ and $1\le\ell\le D_j$. For $c=d$, take
\[
\begin{aligned}
&(hq)_{j,1}^{(1;cc)}(y,\gamma)=\frac{4-D_j}{4}\mathsf s_{j,cc}\widetilde\phi_{j,\lambda_{j,c}}^x(y,\gamma),\\
&(hq)_{j,\ell}^{(2;cc)}(y,\gamma)=\frac14\mathsf s_{j,cc}\widetilde\phi_{j,\lambda_{j,c}}^x(y,\gamma).
\end{aligned}
\]
For $c<d$, take
\[
\begin{aligned}
&(hq)_{j,1}^{(1;cd)}(y,\gamma)=\frac{1+r_{j,cd}}2\mathsf s_{j,cd}\widetilde\phi_{j,\chi_{j,cd}^x}^x(y,\gamma),\\
&(hq)_{j,\ell}^{(2;cd)}(y,\gamma)=-\delta_{\ell c}(hq)_{j,1}^{(1;cd)}(y,\gamma).
\end{aligned}
\]
This gives
\begin{equation}\label{eq:x-space}
\mathscr L_{j,\gamma}^{x,\mathrm{ctrl}}=\left\{X\mapsto\sum_{a,b}\mathsf s_{j,ab}P_{j,a}XP_{j,b}:\mathsf s_{j,ab}=\mathsf s_{j,ba}\in\mathbb C\right\}.
\end{equation}

\subsection{Momentum readout}\label{subsec:momentum-space}

For momentum readouts, only the identity control branch contributes to the sum in Eq.~\eqref{eq:local-space-definition}. The map and its coefficients, with $\omega\in\mathcal S_j^p$, are
\begin{equation}\label{eq:p-weighted-map}
\begin{aligned}
&\mathcal L_{j,\gamma}^p[h](X)=\int h_{j,1}^{(1),p}(y,\gamma)K_{j,\gamma}^p(y)^\dagger XK_{j,\gamma}^p(y)\,dy=\sum_{a,b}\mathsf z_j(\lambda_{j,a}-\lambda_{j,b})P_{j,a}XP_{j,b},\\
&\mathsf z_j(\omega)=\sqrt{2/\pi}\int h_{j,1}^{(1),p}(y,\gamma)e^{-2y^2}e^{i\gamma\omega y}\,dy.
\end{aligned}
\end{equation}
Consequently, its coefficient matrix depends on $(a,b)$ only through $\lambda_{j,a}-\lambda_{j,b}$.

We now show that every function $\mathsf z_j:\mathcal S_j^p\to\mathbb C$ can occur, so that
\[
\mathcal P_j^p=\left\{[\mathsf z_j(\lambda_{j,a}-\lambda_{j,b})]_{a,b=1}^{D_j}:\mathsf z_j:\mathcal S_j^p\to\mathbb C\right\}.
\]

For any prescribed $\mathsf z_j$, use the inverse of the Gram matrix defined in Eq.~\eqref{eq:Gp} to choose
\[
h_{j,1}^{(1),p}(y,\gamma)=\sqrt{2/\pi}e^{-2y^2}\sum_{\omega,\omega'\in\mathcal S_j^p}\mathsf z_j(\omega)(\mathsf G_{j,\gamma}^p)^{-1}(\omega,\omega')e^{-i\gamma\omega'y}.
\]
This is a bounded function at fixed $\gamma$. Substituting it into Eq.~\eqref{eq:p-weighted-map} and using the Gram-matrix identity gives the prescribed values of $\mathsf z_j$ at every $\omega\in\mathcal S_j^p$. Therefore,
\begin{equation}\label{eq:p-space}
\mathscr L_{j,\gamma}^{p,\mathrm{ctrl}}=\mathscr L_{j,\gamma}^p=\left\{X\mapsto\sum_{a,b}\mathsf z_j(\lambda_{j,a}-\lambda_{j,b})P_{j,a}XP_{j,b}:\mathsf z_j:\mathcal S_j^p\to\mathbb C\right\}.
\end{equation}

\subsection{Space of estimable operators}

We now combine the local coefficient spaces $\mathcal P_j^x$ and $\mathcal P_j^p$ to determine which operators $G_{\boldsymbol\eta,\mathbf U}$ the complete protocol can estimate. Let $\mathbf I$ denote the control sequence in which every control operation is the identity. When the local maps are composed, each entry of the resulting coefficient tensor is a product of entries of the local coefficient matrices. Finite sums of such products, with $\mathsf s_{r,j}\in\mathcal P_j^{\eta_j}$, have the form
\[
S_{\mathbf a\mathbf b}=\sum_r\prod_{j=0}^{N-1}(\mathsf s_{r,j})_{a_jb_j},
\]
that is, $S\in\bigotimes_{j=0}^{N-1}\mathcal P_j^{\eta_j}$. The same coefficient tensor $S$, independent of $\rho_0$ and $\mathbf U$, determines $G_{\boldsymbol\eta,\mathbf U}$ for every $\mathbf U$. The space of estimable operators is therefore
\begin{equation}\label{eq:controlled-space}
\mathcal C_\gamma^{\boldsymbol\eta,\mathrm{ctrl}}=\mathcal C^{\boldsymbol\eta,\mathrm{ctrl}}
=\left\{\sum_{\mathbf a,\mathbf b}S_{\mathbf a\mathbf b}
(Q_{\mathbf a,\mathbf U}^{\mathbf I})^\dagger Q_{\mathbf b,\mathbf U}^{\mathbf I}
\;\middle|\;S\in\bigotimes_{j=0}^{N-1}\mathcal P_j^{\eta_j}\right\}.
\end{equation}
Neither the coefficient spaces $\mathcal P_j^{\eta_j}$ nor the operators $Q_{\mathbf a,\mathbf U}^{\mathbf I}$ depend on $\gamma$. Equation~\eqref{eq:controlled-space} therefore proves the equality in Eq.~\eqref{eq:strength}.

\section{Proof of Theorem 1}\label{sec:theorem}

Let $\mathcal H$ be the system Hilbert space. We assume that every $A_j$ has at least two distinct eigenvalues and that $\eta_{N-1}=x$, as in the main text.

Let $P_{j,a}(t_j)$ denote the spectral projector of $A_j(t_j)$ associated with $\lambda_{j,a}$. Since the initial state is arbitrary and the unitary evolutions between successive measurements $\mathbf U=(U_{1:0},\ldots,U_{N-1:N-2})$ can be chosen independently, the universal unbiasedness condition is
\begin{equation}\label{eq:paths}
\begin{aligned}
&\int f_{\boldsymbol\eta}(\mathbf y,\gamma)E_{\gamma,\mathbf U}^{\boldsymbol\eta}(\mathbf y)\,d\mathbf y
-G_{\boldsymbol\eta,\mathbf U}\\
&=\sum_{\mathbf a,\mathbf a'}\mathsf W_{\boldsymbol\eta}(\mathbf a,\mathbf a')
P_{0,a_0}(t_0)\cdots P_{N-1,a_{N-1}}(t_{N-1})
P_{N-1,a'_{N-1}}(t_{N-1})\cdots P_{0,a'_0}(t_0)=0,
\end{aligned}
\end{equation}
where
\begin{equation}\label{eq:W}
\mathsf W_{\boldsymbol\eta}(\mathbf a,\mathbf a')
=\int f_{\boldsymbol\eta}(\mathbf y,\gamma)\prod_{j=0}^{N-1}
k_{j,\gamma}^{\eta_j}(y_j;a_j,a'_j)\,d\mathbf y
-\prod_{j=0}^{N-1}\chi_{j,a_ja'_j}^{\eta_j}.
\end{equation}

\subsection{Necessity}

When $\dim\mathcal H=2$, every nonscalar observable has exactly two distinct eigenvalues, so the spectral condition in the theorem holds automatically and necessity is trivial. We therefore only need to consider $\dim\mathcal H\ge3$ below.

We first solve Eq.~\eqref{eq:paths} for a coefficient array $\mathsf W_{\boldsymbol\eta}$ independent of $\mathbf U$. We prove that the weighted sum of products of projectors in this equation vanishes for every $\mathbf U$ if and only if the coefficients have the form
\begin{equation}\label{eq:path-kernel}
\mathsf W_{\boldsymbol\eta}(\mathbf a,\mathbf a')
=\sum_{j=0}^{N-1}(1-\delta_{a_j,a'_j})
g_j(a_0,a'_0,\ldots,a_j,a'_j).
\end{equation}
Here each $g_j$ is an arbitrary complex-valued function of the pairs of spectral indices $(a_k,a'_k)$ at times $k=0,\ldots,j$ and does not depend on later spectral indices.

First, direct substitution of Eq.~\eqref{eq:path-kernel} into the left-hand side of Eq.~\eqref{eq:paths} gives zero for every $\mathbf U$, so every coefficient array of this form is a solution.

Next, take any coefficient array satisfying Eq.~\eqref{eq:paths} for every $\mathbf U$. To show that it has the form in Eq.~\eqref{eq:path-kernel}, we first consider the projectors at a single measurement time. Let $\{P_a\}$ be a finite orthogonal resolution of the identity consisting of at least two nonzero projectors, and let $X_{ab}$ be fixed bounded operators independent of the unitary $U$. If for every $U$ there is a scalar $s(U)$ such that
\[
\sum_{a,b}P_aU^\dagger X_{ab}UP_b=s(U)I,
\]
then every $X_{ab}$ is a scalar multiple of the identity.

To prove this statement, take $a\ne b$ and fix unit vectors $|u\rangle\in\operatorname{ran}P_a$ and $|v\rangle\in\operatorname{ran}P_b$. For any orthonormal pair $|x\rangle,|y\rangle$, there is a unitary $U$ such that $U|u\rangle=|x\rangle$ and $U|v\rangle=|y\rangle$. The off-diagonal matrix element gives $\langle x|X_{ab}|y\rangle=0$ whenever $\langle x|y\rangle=0$, so $X_{ab}|y\rangle$ is proportional to $|y\rangle$ for every $|y\rangle$. Thus $X_{ab}$ is scalar when $a\ne b$.

For the diagonal terms, fix $a$ and choose any $b\ne a$. The two diagonal matrix elements give
\[
\langle x|X_{aa}|x\rangle=s(U)=\langle y|X_{bb}|y\rangle.
\]
Since $\dim\mathcal H\ge3$, choose unitaries $U,U'$ satisfying $\langle u|U'^\dagger U|u\rangle=0$ and $\langle v|U'^\dagger U|v\rangle=1$.
Their images $|x\rangle=U|u\rangle$ and $|x'\rangle=U'|u\rangle$ are orthogonal, while $|y\rangle=U|v\rangle=U'|v\rangle$. The diagonal equalities above hold for every such choice and give
\[
\langle x|X_{aa}|x\rangle=s(U)=\langle y|X_{bb}|y\rangle
=s(U')=\langle x'|X_{aa}|x'\rangle.
\]
For fixed $|y\rangle$, these choices cover every orthonormal pair in its orthogonal complement, and $s(U)=\langle y|X_{bb}|y\rangle$ has the same value for all of them. The complex polarization identity therefore gives
\[
(I-|y\rangle\langle y|)X_{aa}(I-|y\rangle\langle y|)=s(U)(I-|y\rangle\langle y|).
\]
This holds for every $|y\rangle$. Any two such orthogonal complements have a nonzero intersection, so their scalar values coincide. Hence $X_{aa}=s(U)I$ is scalar, proving the statement.

We now apply this statement to any coefficient array satisfying Eq.~\eqref{eq:paths}. For $N\ge2$, grouping the terms by the first pair of indices $(a_0,a'_0)$ gives, for every $\mathbf U$,
\[
\sum_{a_0,a'_0}P_{0,a_0}U_{1:0}^\dagger
X^{(1)}_{a_0;a'_0}U_{1:0}P_{0,a'_0}=0.
\]
The operators at later steps are defined recursively, for $1\le j<N-1$, by
\[
X^{(j)}_{a_{j-1},\ldots,a_0;\,a'_{j-1},\ldots,a'_0}
=\sum_{a_j,a'_j}P_{j,a_j}U_{j+1:j}^\dagger
X^{(j+1)}_{a_j,\ldots,a_0;\,a'_j,\ldots,a'_0}
U_{j+1:j}P_{j,a'_j},
\]
with the terminal condition
\[
X^{(N-1)}_{a_{N-2},\ldots,a_0;\,a'_{N-2},\ldots,a'_0}
=\sum_{a_{N-1},a'_{N-1}}\delta_{a_{N-1},a'_{N-1}}
\mathsf W_{\boldsymbol\eta}(\mathbf a,\mathbf a')P_{N-1,a_{N-1}}.
\]
Each $X^{(j)}$ is independent of $U_{j:j-1}$. Applying the result above first to the equation for $X^{(1)}$ and then at each step of the recursion shows that every $X^{(j)}$ is scalar and that $X^{(1)}_{a_0;a_0}=0$. In each recursion, every $X^{(j+1)}$ with $a_j=a'_j$ equals $X^{(j)}$. Taking $a_k=a'_k$ at every later step shows that these scalars are independent of the subsequent unitary evolutions.

The first step and the successive differences for $1\le j<N-1$ can thus be written as
\[
\begin{aligned}
&X^{(1)}_{a_0;a'_0}
=(1-\delta_{a_0,a'_0})g_0(a_0,a'_0)I,\\
&X^{(j+1)}_{a_j,\ldots,a_0;\,a'_j,\ldots,a'_0}
=X^{(j)}_{a_{j-1},\ldots,a_0;\,a'_{j-1},\ldots,a'_0}
+(1-\delta_{a_j,a'_j})g_j(a_0,a'_0,\ldots,a_j,a'_j)I.
\end{aligned}
\]
Each $g_j$ depends only on the index pairs at times $0,\ldots,j$. For $a_{N-1}=a'_{N-1}$, combining these recurrence relations with the terminal condition gives
\[
\mathsf W_{\boldsymbol\eta}(\mathbf a,\mathbf a')
=\sum_{j=0}^{N-2}(1-\delta_{a_j,a'_j})g_j(a_0,a'_0,\ldots,a_j,a'_j).
\]
For $a_{N-1}\ne a'_{N-1}$, define $g_{N-1}$ as the difference between $\mathsf W_{\boldsymbol\eta}(\mathbf a,\mathbf a')$ and this sum. The two cases for the final indices together give Eq.~\eqref{eq:path-kernel}.

We now apply this general solution to the coefficients $\mathsf W_{\boldsymbol\eta}$ in Eq.~\eqref{eq:W}. Suppose that a universally unbiased estimator exists. At the final measurement, choose indices $c,d$ with $\lambda_{N-1,c}\ne\lambda_{N-1,d}$. With the earlier index pairs fixed, Eq.~\eqref{eq:path-kernel} takes the same value for the final index pairs $(c,c)$ and $(d,d)$ because its $j=N-1$ term vanishes at both pairs and its other terms do not depend on the terminal indices. Subtracting the corresponding expressions in Eq.~\eqref{eq:W} gives
\begin{equation}\label{eq:terminal-contrast}
\begin{aligned}
&\int f_{\boldsymbol\eta}(\mathbf y,\gamma)
\prod_{j=0}^{N-2}k_{j,\gamma}^{\eta_j}(y_j;a_j,a'_j)
\big[k_{N-1,\gamma}^x(y_{N-1};c,c)
-k_{N-1,\gamma}^x(y_{N-1};d,d)\big]\,d\mathbf y\\
&=(\lambda_{N-1,c}-\lambda_{N-1,d})
\prod_{j=0}^{N-2}\chi_{j,a_ja'_j}^{\eta_j}.
\end{aligned}
\end{equation}

Fix a time $0\le j<N-1$ with $\eta_j=x$ and two index pairs $(a,a')$, $(b,b')$ satisfying $\chi_{j,aa'}^x=\chi_{j,bb'}^x\ne0$. Since every $A_k$ is nonscalar, we can choose the remaining pairs $\{(a_k,a'_k):0\le k<N-1,\ k\ne j\}$ such that
\[\prod_{\substack{0\le k<N-1\\k\ne j}}\chi_{k,a_ka'_k}^{\eta_k}\ne0.\]
Keep the chosen pairs $(a_k,a'_k)$ with $k\ne j,N-1$ fixed in the following four evaluations of $\mathsf W_{\boldsymbol\eta}$. Equation~\eqref{eq:path-kernel} requires
\[
\begin{aligned}
&\mathsf W_{\boldsymbol\eta}\big((a_0,\ldots,a,\ldots,a_{N-2},c),(a'_0,\ldots,a',\ldots,a'_{N-2},c)\big)\\
&-\mathsf W_{\boldsymbol\eta}\big((a_0,\ldots,a,\ldots,a_{N-2},d),(a'_0,\ldots,a',\ldots,a'_{N-2},d)\big)=0,\\
&\mathsf W_{\boldsymbol\eta}\big((a_0,\ldots,b,\ldots,a_{N-2},c),(a'_0,\ldots,b',\ldots,a'_{N-2},c)\big)\\
&-\mathsf W_{\boldsymbol\eta}\big((a_0,\ldots,b,\ldots,a_{N-2},d),(a'_0,\ldots,b',\ldots,a'_{N-2},d)\big)=0.
\end{aligned}
\]
The entries $a,b$ occupy the time-$j$ coordinate of the first argument, and $a',b'$ occupy that of the second argument. Substituting Eq.~\eqref{eq:W} into these two equalities gives Eq.~\eqref{eq:terminal-contrast} for $(a,a')$ and $(b,b')$, respectively. The corresponding position kernels satisfy
\[k_{j,\gamma}^x(y;a,a')=\frac{1+r_{j,bb'}}{1+r_{j,aa'}}k_{j,\gamma}^x(y;b,b').\]
Subtract $(1+r_{j,bb'})/(1+r_{j,aa'})$ times the equation for $(b,b')$ from the equation for $(a,a')$. The two integral terms cancel, so
\[
0=(\lambda_{N-1,c}-\lambda_{N-1,d})\left[\chi_{j,aa'}^x-\frac{1+r_{j,bb'}}{1+r_{j,aa'}}\chi_{j,bb'}^x\right]\prod_{\substack{0\le k<N-1\\k\ne j}}\chi_{k,a_ka'_k}^{\eta_k}.
\]
By the choice of $c,d$ and of the indices at the other measurement times, every factor outside the square brackets is nonzero. Since $\chi_{j,aa'}^x=\chi_{j,bb'}^x\ne0$, this equality requires $r_{j,aa'}=r_{j,bb'}$. By the definition of $r_{j,ab}$, the necessary spectral condition is
\begin{equation}\label{eq:spectral-necessity}
(\lambda_{j,a}-\lambda_{j,a'})^2=(\lambda_{j,b}-\lambda_{j,b'})^2.
\end{equation}

If instead $\chi_{j,aa'}^x=\chi_{j,bb'}^x=0$, the right-hand sides of both instances of Eq.~\eqref{eq:terminal-contrast} vanish. Their comparison therefore imposes no spectral condition.

\subsection{Sufficiency}

Fix $0<\gamma<\infty$ and assume that every nonfinal position readout satisfies the spectral condition in Theorem~1 of the main text. The following construction applies to every $\dim\mathcal H\ge2$. Let $\mathbf y_x=(y_j)_{j:\eta_j=x}$ collect all position readouts. In this sufficiency proof, we choose the estimator in the form
\[
f_{\boldsymbol\eta}(\mathbf y,\gamma)
=F_x(\mathbf y_x,\gamma)\prod_{j:\eta_j=p}h_{j,1}^{(1),p}(y_j,\gamma).
\]
At each momentum readout, use the construction in Section~\ref{sec:space} with $\mathsf z_j(\omega)=-i\omega/2$. For any spectrum, this supplies a bounded local factor satisfying, for every $\omega\in\mathcal S_j^p$,
\begin{equation}\label{eq:h-constraints}
\int h_{j,1}^{(1),p}(y,\gamma)e^{i\gamma\omega y}\,d\mu_p(y)=-i\omega/2,
\end{equation}
where $d\mu_p(y)=\sqrt{2/\pi}e^{-2y^2}dy$. For all $1\le a,b\le D_j$, it satisfies
\[
\int h_{j,1}^{(1),p}(y,\gamma)k_{j,\gamma}^p(y;a,b)\,dy=\chi_{j,ab}^p.
\]
With these factors chosen, Eq.~\eqref{eq:W} becomes
\[
\mathsf W_{\boldsymbol\eta}(\mathbf a,\mathbf a')
=\prod_{j:\eta_j=p}\chi_{j,a_ja'_j}^p
\left[\int F_x(\mathbf y_x,\gamma)\prod_{j:\eta_j=x}k_{j,\gamma}^x(y_j;a_j,a'_j)\,d\mathbf y_x
-\prod_{j:\eta_j=x}\chi_{j,a_ja'_j}^x\right].
\]

For the position readouts we choose
\[
F_x(\mathbf y_x,\gamma)
=y_{N-1}\prod_{\substack{j<N-1\\\eta_j=x}}h_{j,1}^{(1),x}(y_j,\gamma).
\]
We construct the local factors $h_{j,1}^{(1),x}$ independently of $\rho_0$ and $\mathbf U$.

Fix $j<N-1$ with $\eta_j=x$. The spectral condition implies that, for every $1\le a\le b\le D_j$ with $\chi_{j,ab}^x\ne0$, there is no pair $1\le c\le d\le D_j$ with $(c,d)\ne(a,b)$ and $\chi_{j,cd}^x=\chi_{j,ab}^x$.

Use the set $\mathcal S_j^x$ and positive definite Gram matrix $\mathsf G_{j,\gamma}^x$ from Eqs.~\eqref{eq:spectral-sets} and \eqref{eq:Gx}. For $\vartheta,\vartheta'\in\mathcal S_j^x$, the real bounded functions $\widetilde\phi_{j,\vartheta}^x$ defined in Eq.~\eqref{eq:pair} satisfy
\[
\int\widetilde\phi_{j,\vartheta}^x(y,\gamma)\phi_{\vartheta',\gamma}(y)\,dy=\delta_{\vartheta,\vartheta'}.
\]
For $j<N-1$ with $\eta_j=x$, define the local estimator explicitly by
\[
h_{j,1}^{(1),x}(y,\gamma)=\sum_{1\le a\le b\le D_j}(1+r_{j,ab})\chi_{j,ab}^x\widetilde\phi_{j,\chi_{j,ab}^x}^x(y,\gamma).
\]
For any $1\le c\le d\le D_j$, substitution gives
\[
\int h_{j,1}^{(1),x}(y,\gamma)\phi_{\chi_{j,cd}^x,\gamma}(y)\,dy=\sum_{\substack{1\le a\le b\le D_j\\\chi_{j,ab}^x=\chi_{j,cd}^x}}(1+r_{j,ab})\chi_{j,ab}^x.
\]
If $\chi_{j,cd}^x\ne0$, the spectral condition leaves only the term $(a,b)=(c,d)$ in this sum. If $\chi_{j,cd}^x=0$, every summand is zero. In both cases the integral equals $(1+r_{j,cd})\chi_{j,cd}^x$. Equation~\eqref{eq:instrument-expansion} therefore yields
\[
\int h_{j,1}^{(1),x}(y,\gamma)k_{j,\gamma}^x(y;c,d)\,dy
=\frac{1}{1+r_{j,cd}}\int h_{j,1}^{(1),x}(y,\gamma)\phi_{\chi_{j,cd}^x,\gamma}(y)\,dy
=\chi_{j,cd}^x.
\]
Symmetry under $c\leftrightarrow d$ extends this identity to all $1\le c,d\le D_j$. Thus the local factors satisfy the moment conditions for all position readouts before the final measurement. Using the final position readout $y_{N-1}$ itself as the last factor gives a universally unbiased estimator.

\section{Bias and variance of factorized estimators}\label{sec:expansions}

Let $\|A_j\|=\max_a|\lambda_{j,a}|$ be the operator norm. Recall the factorized estimator used for the conventional protocol in the main text:
\[f_{\boldsymbol\eta}(\mathbf y,\gamma)=\prod_{j:\eta_j=x}y_j\prod_{j:\eta_j=p}h_{j,1}^{(1),p}(y_j,\gamma).\]
As follows, we use $\operatorname{Var}f=\mathrm E|f-\mathrm Ef|^2$ to denote the variances for complex-valued estimators,.

\subsection{Bias expansion}

Gaussian integration gives the first-moment map for a position readout in terms of $C_A^x$ and $C_A^p$, as defined in the main text,
\begin{equation}\label{eq:first-x}
\int yK_{j,\gamma}^x(y)^\dagger XK_{j,\gamma}^x(y)\,dy=\exp\!\left[\frac{\gamma^2}{2}(C_{A_j}^p)^2\right]\!\big(C_{A_j}^x(X)\big).
\end{equation}
At momentum readouts, Eq.~\eqref{eq:h-constraints} gives $C_{A_j}^p$ exactly. Thus only position readouts contribute to the bias.

Let $\boldsymbol\ell=(\ell_0,\ldots,\ell_{N-1})$ contain nonnegative integers, with $\ell_j=0$ whenever $\eta_j=p$, and put $|\boldsymbol\ell|=\sum_j\ell_j$. Expanding the position maps and composing them in time order gives the bias series $\epsilon(\gamma)=\sum_{k\ge1}b_k\gamma^{2k}$ with coefficients
\begin{equation}\label{eq:bias}
b_k=\frac1{2^k}\sum_{|\boldsymbol\ell|=k}\frac{\operatorname{Tr}\!\left[\rho_0\left(\mathop{\bigcirc}_{j=0}^{N-1}\big[(C_{A_j(t_j)}^p)^{2\ell_j}\circ C_{A_j(t_j)}^{\eta_j}\big]\right)(I)\right]}{\prod_{j=0}^{N-1}\ell_j!}.
\end{equation}
The composition has increasing $j$ from left to right, so the latest map acts first. All $b_k$ are real.

If at least one $b_k$ is nonzero, define $n=\min\{2k:b_k\ne0\}$. The leading squared bias is
\begin{equation}\label{eq:bias-squared}
\epsilon(\gamma)^2=b_{n/2}^{2}\gamma^{2n}[1+O(\gamma^2)].
\end{equation}
If every $b_k$ is zero, set $n=\infty$; the bias then vanishes at every strength. In particular, when no nonfinal readout is a position readout, $C_{A_{N-1}}^x(I)=A_{N-1}$, so the estimator is unbiased.

\subsection{Second moments of estimators for momentum readouts}\label{subsec:momentum-variance}

We first derive a lower bound on the second moment of any complex-valued local factor satisfying Eq.~\eqref{eq:h-constraints}, then construct a polynomial estimator whose second moment scales as $\gamma^{-2}$ as $\gamma\to0$. Subtracting the equations at $\omega$ and $-\omega$ gives $\int h_{j,1}^{(1),p}(y,\gamma)\sin(\gamma\omega y)\,d\mu_p(y)=-\omega/2$. Cauchy--Schwarz therefore implies, for every nonzero $\omega\in\mathcal S_j^p$,
\[
\gamma^2\int|h_{j,1}^{(1),p}(y,\gamma)|^2\,d\mu_p(y)
\ge\frac{\gamma^2\omega^2}{2(1-e^{-\gamma^2\omega^2/2})}.
\]

Let $\omega_{j,1}<\cdots<\omega_{j,m_j}$ be all the positive elements of $\mathcal S_j^p$, with $m_j=(|\mathcal S_j^p|-1)/2$. For each $\gamma>0$, define $d_{j,r}(\gamma)$ for $0\le r<m_j$ by the equations indexed by $1\le\ell\le m_j$,
\begin{equation}\label{eq:momentum-polynomial-coefficients}
\sum_{r=0}^{m_j-1}d_{j,r}(\gamma)(\gamma^2\omega_{j,\ell}^2)^r=e^{\gamma^2\omega_{j,\ell}^2/8}.
\end{equation}
The distinct values of $\omega_{j,\ell}^2$ make the coefficient matrix invertible. Choose
\[
h_{j,1}^{(1),p}(y,\gamma)=\frac{e^{2y^2}}{2\gamma}\sum_{r=0}^{m_j-1}(-1)^r d_{j,r}(\gamma)\frac{d^{2r+1}}{dy^{2r+1}}e^{-2y^2}.
\]
This estimator is an odd polynomial of degree at most $2m_j-1$ and has a finite second moment with respect to $d\mu_p$.

To verify the momentum-readout condition, we integrate by parts repeatedly and evaluate the Gaussian Fourier integral to obtain
\[
\sqrt{\frac2\pi}\int\frac{d^r e^{-2y^2}}{dy^r}e^{i\gamma\omega y}\,dy
=(-i\gamma\omega)^r\int e^{i\gamma\omega y}\,d\mu_p(y)
=(-i\gamma\omega)^r e^{-\gamma^2\omega^2/8}.
\]
Therefore, Eq.~\eqref{eq:momentum-polynomial-coefficients} gives, for $\omega\in\mathcal S_j^p$,
\[
\begin{aligned}
&\int h_{j,1}^{(1),p}(y,\gamma)e^{i\gamma\omega y}\,d\mu_p(y)
=\frac{e^{-\gamma^2\omega^2/8}}{2\gamma}
\sum_{r=0}^{m_j-1}(-1)^rd_{j,r}(\gamma)(-i\gamma\omega)^{2r+1}\\
&=-\frac{i\omega}{2}e^{-\gamma^2\omega^2/8}
\sum_{r=0}^{m_j-1}d_{j,r}(\gamma)(\gamma^2\omega^2)^r
=-\frac{i\omega}{2}.
\end{aligned}
\]

To expand $\int|h_{j,1}^{(1),p}(y,\gamma)|^2\,d\mu_p(y)$ for small $\gamma$, we first determine the expansion of $d_{j,r}(\gamma)$. Expanding the exponential on the right-hand side of Eq.~\eqref{eq:momentum-polynomial-coefficients} gives, for $1\le\ell\le m_j$,
\[
\sum_{r=0}^{m_j-1}\gamma^{2r}d_{j,r}(\gamma)\omega_{j,\ell}^{2r}
=\sum_{r=0}^{m_j-1}\frac{\gamma^{2r}\omega_{j,\ell}^{2r}}{8^rr!}+O(\gamma^{2m_j}).
\]
The fixed matrix with entries $\omega_{j,\ell}^{2r}$ is invertible. Solving for the coefficient differences yields, for $0\le r<m_j$,
\[
d_{j,r}(\gamma)=\frac1{8^rr!}+O\!\left(\gamma^{2(m_j-r)}\right).
\]

The polynomial $e^{2y^2}d^ke^{-2y^2}/dy^k$ has degree $k$ and leading coefficient $(-4)^k$. For nonnegative integers $r\ge k$, integrating by parts $r$ times therefore gives
\[
\begin{aligned}
&\sqrt{\frac2\pi}\int e^{2y^2}\frac{d^r e^{-2y^2}}{dy^r}\frac{d^k e^{-2y^2}}{dy^k}\,dy
=(-1)^r\sqrt{\frac2\pi}\int e^{-2y^2}\frac{d^r}{dy^r}\!\left(e^{2y^2}\frac{d^k e^{-2y^2}}{dy^k}\right)dy\\
&=4^r r!\,\delta_{rk}.
\end{aligned}
\]
The integral is symmetric in $r$ and $k$, so the identity also holds for $r<k$. Substitution of the polynomial estimator gives
\[
\gamma^2\int|h_{j,1}^{(1),p}(y,\gamma)|^2\,d\mu_p(y)=\sum_{r=0}^{m_j-1}16^r(2r+1)!\,d_{j,r}(\gamma)^2.
\]
Substituting the coefficient expansion gives
\begin{equation}\label{eq:momentum-variance-scaling}
\begin{aligned}
&\int|h_{j,1}^{(1),p}(y,\gamma)|^2\,d\mu_p(y)=c_j\gamma^{-2}[1+O(\gamma^2)],\\
&c_j=\sum_{r=0}^{m_j-1}\frac{(2r+1)!}{4^r(r!)^2}.
\end{aligned}
\end{equation}
Here $1\le c_j<\infty$.

\subsection{Variance in the weak-measurement limit}

We now compute the variance using the local estimators just constructed for momentum readouts. The second moments for the position and momentum readouts satisfy
\begin{equation}\label{eq:second}
\begin{aligned}
&\int y^2K_{j,\gamma}^x(y)^\dagger K_{j,\gamma}^x(y)\,dy=\gamma^{-2}I+A_j^2,\\
&\int|h_{j,1}^{(1),p}(y,\gamma)|^2K_{j,\gamma}^p(y)^\dagger K_{j,\gamma}^p(y)\,dy
=\left(\int|h_{j,1}^{(1),p}|^2\,d\mu_p\right)I.
\end{aligned}
\end{equation}
Since $0\le A_j^2\le\|A_j\|^2I$ and the second-moment maps are positive, successive integration gives, for every initial state and free evolution,
\begin{equation}\label{eq:second-full}
\begin{aligned}
&\mathrm E|f_{\boldsymbol\eta}|^2
=\left(\prod_{j:\eta_j=x}\gamma^{-2}\right)
\left(\prod_{j:\eta_j=p}\int|h_{j,1}^{(1),p}|^2\,d\mu_p\right)[1+O(\gamma^2)]\\
&=\left(\prod_{j:\eta_j=p}c_j\right)\gamma^{-2N}[1+O(\gamma^2)].
\end{aligned}
\end{equation}
The mean remains bounded as $\gamma\to0$. Therefore,
\begin{equation}\label{eq:variance}
\operatorname{Var}f_{\boldsymbol\eta}
=\left(\prod_{j:\eta_j=p}c_j\right)\gamma^{-2N}[1+O(\gamma^2)].
\end{equation}

\section{Optimization of factorized estimators}
\label{sec:optimization}

We optimize the control probabilities, local estimators, and common measurement strength within the factorized class defined in the main text. The optimization uses only the measured spectra and the instrument parameters. For a given control sequence, the estimator is the product of the local factors,
\[
f_{\mathbf T_{\boldsymbol\beta}^{(\boldsymbol\alpha),\boldsymbol\eta}}(\mathbf y,\gamma)=\prod_{j=0}^{N-1}h_{j,\beta_j}^{(\alpha_j),\eta_j}(y_j,\gamma).
\]

\subsection{Local second moments and the minimax variance}

Summing over control sequences and integrating the readouts gives the second moment of the factorized estimator,
\[
\mathrm E_{\rho_0,\mathbf U}\left|f_{\mathbf T_{\boldsymbol\beta}^{(\boldsymbol\alpha),\boldsymbol\eta}}\right|^2
=\sum_{\boldsymbol\alpha,\boldsymbol\beta}\int
\left|f_{\mathbf T_{\boldsymbol\beta}^{(\boldsymbol\alpha),\boldsymbol\eta}}(\mathbf y,\gamma)\right|^2
\operatorname{Tr}\!\left[\rho_0E_{\gamma,\mathbf U}^{\boldsymbol\eta,\mathrm{ctrl}}\bigl(\mathbf y,\mathbf T_{\boldsymbol\beta}^{(\boldsymbol\alpha),\boldsymbol\eta}\bigr)\right]\,d\mathbf y.
\]
Using the local maps in Eq.~\eqref{eq:record-map}, we collect each local sum and integral into a single map for either readout type:
\[
\mathcal R_j^{\eta_j}(X)=\sum_{\alpha,\beta}q_{j,\beta}^{(\alpha)}
\int\left|h_{j,\beta}^{(\alpha),\eta_j}(y,\gamma)\right|^2
\mathcal K_{j,\gamma,\beta}^{(\alpha),\eta_j}(y)(X)\,dy.
\]
For the position and momentum readouts, define the local second moments by
\begin{equation}\label{eq:local-cost}
\begin{aligned}
&\sigma_{j,a}^x=\sum_{\alpha,\beta}q_{j,\beta}^{(\alpha)}
\int\left|h_{j,\beta}^{(\alpha),x}(y,\gamma)\right|^2
\phi_{\lambda_{j,a},\gamma}(y)\,dy,\\
&\sigma_j^p=\int\left|h_{j,1}^{(1),p}(y,\gamma)\right|^2\,d\mu_p(y).
\end{aligned}
\end{equation}
For position readouts, set $\sigma_j^x=\max_a\sigma_{j,a}^x$. The spectral decompositions of the measurement operators give $\mathcal R_j^x(I)=\sum_a\sigma_{j,a}^xP_{j,a}$ and $\mathcal R_j^p(I)=\sigma_j^pI$. Thus the largest eigenvalue of $\mathcal R_j^{\eta_j}(I)$ is $\sigma_j^{\eta_j}$.

Set $X_{N-1}=\mathcal R_{N-1}^{x}(I)$ and recurse over $j=N-2,\ldots,0$ using
\[
X_j=\mathcal R_j^{\eta_j}\!\left(U_{j+1:j}^{\dagger}X_{j+1}U_{j+1:j}\right).
\]
The second moment of the factorized estimator can therefore be written as
\[
\begin{aligned}
&\mathrm E_{\rho_0,\mathbf U}\left|f_{\mathbf T_{\boldsymbol\beta}^{(\boldsymbol\alpha),\boldsymbol\eta}}\right|^2
=\operatorname{Tr}\!\left[\rho_0\mathcal R_0^{\eta_0}
\left(\mathcal U_{1:0}\left(\cdots\mathcal R_{N-1}^{x}(I)\cdots\right)\right)\right]\\
&=\operatorname{Tr}(\rho_0X_0).
\end{aligned}
\]

For any positive semidefinite $X$, the nonnegative weights in the defining integral imply
\[
0\le\langle v|\mathcal R_j^{\eta_j}(X)|v\rangle
\le \|X\|\langle v|\mathcal R_j^{\eta_j}(I)|v\rangle
\le \|X\|\sigma_j^{\eta_j}\langle v|v\rangle.
\]
Using $\|U_{j+1:j}v\|=\|v\|$ at each step of the recursion gives, for $j=N-1,\ldots,0$,
\[
0\le\langle v|X_j|v\rangle
\le\left(\prod_{k=j}^{N-1}\sigma_k^{\eta_k}\right)\langle v|v\rangle.
\]
Taking the trace against any normalized density operator $\rho_0$ gives, for every initial state and free evolution,
\begin{equation}\label{eq:worst-bound}
\operatorname{Var}_{\rho_0,\mathbf U}f_{\mathbf T_{\boldsymbol\beta}^{(\boldsymbol\alpha),\boldsymbol\eta}}
\le\mathrm E_{\rho_0,\mathbf U}\left|f_{\mathbf T_{\boldsymbol\beta}^{(\boldsymbol\alpha),\boldsymbol\eta}}\right|^2
\le\prod_{j=0}^{N-1}\sigma_j^{\eta_j}.
\end{equation}

We next prove that, for every fixed measurement strength, local control probabilities, and estimators,
\[
\max_{\rho_0,\mathbf U}\mathrm E_{\rho_0,\mathbf U}\left|f_{\mathbf T_{\boldsymbol\beta}^{(\boldsymbol\alpha),\boldsymbol\eta}}\right|^2
=\prod_{j=0}^{N-1}\sigma_j^{\eta_j}.
\]

At each position readout choose $a_{j,*}$ maximizing $\sigma_{j,a}^x$; at each momentum readout choose any spectral index. For each $j$, choose a vector $|e_j\rangle$ satisfying $P_{j,a_{j,*}}|e_j\rangle=|e_j\rangle$ and $\langle e_j|e_j\rangle=1$.

Extend each $|e_j\rangle$ to an orthonormal basis $\{|e_{j,r}\rangle\}_r$ of the same system Hilbert space, using the same index set and $|e_{j,1}\rangle=|e_j\rangle$. Choose $\rho_0=|e_0\rangle\langle e_0|$ and, for $0\le j<N-1$, set
\[
U_{j+1:j}=\sum_r|e_{j+1,r}\rangle\langle e_{j,r}|.
\]
This choice satisfies $U_{j+1:j}^{\dagger}U_{j+1:j}=U_{j+1:j}U_{j+1:j}^{\dagger}=I$ and $U_{j+1:j}|e_j\rangle=|e_{j+1}\rangle$. Thus the selected eigenvectors may belong to different observables $A_j$.

Since the measurement and spectral control operators act as scalars on $|e_j\rangle$, their spectral decompositions give, for every system operator $X$,
\[
\langle e_j|\mathcal R_j^{\eta_j}(X)|e_j\rangle
=\sigma_j^{\eta_j}\langle e_j|X|e_j\rangle.
\]
For $j<N-1$, applying this identity to the recursion yields
\[
\langle e_j|X_j|e_j\rangle
=\sigma_j^{\eta_j}\langle e_{j+1}|X_{j+1}|e_{j+1}\rangle.
\]
Iterating from $\langle e_{N-1}|X_{N-1}|e_{N-1}\rangle=\sigma_{N-1}^{x}$ gives
\[
\mathrm E_{\rho_0,\mathbf U}\left|f_{\mathbf T_{\boldsymbol\beta}^{(\boldsymbol\alpha),\boldsymbol\eta}}\right|^2
=\langle e_0|X_0|e_0\rangle=\prod_{j=0}^{N-1}\sigma_j^{\eta_j}.
\]
This construction attains the upper bound in Eq.~\eqref{eq:worst-bound} and hence gives the maximal second moment. To bound the variance, we must also account for the mean.

\textit{Factorization of the minimax variance.} Suppose the final measurement uses probe position readout and the spectrum of $A_{N-1}$ is symmetric about zero, with $\lambda_{N-1,\bar b}=-\lambda_{N-1,b}$. As in the main text, $\sigma_{j,*}^{\eta_j}(\gamma)$ is the minimum of $\sigma_j^{\eta_j}$ over all feasible local probabilities and estimators at fixed $\gamma$. Then, at every finite $\gamma>0$,
\begin{equation}\label{eq:exact-worst}
\min_{\mathbf q,h}\max_{\rho_0,\mathbf U}\operatorname{Var}f_{\mathbf T_{\boldsymbol\beta}^{(\boldsymbol\alpha),\boldsymbol\eta}}=\prod_{j=0}^{N-1}\sigma_{j,*}^{\eta_j}(\gamma).
\end{equation}

\textit{Proof.} Fix any feasible local control probabilities and estimators at the given strength. The complete control-sequence probability remains $\prod_{j=0}^{N-1}q_{j,\beta_j}^{(\alpha_j)}$, and the estimator remains the product of its local factors. In particular, the terminal factor is not required to be odd. We first obtain a variance lower bound for this arbitrary fixed choice.

For $N\ge2$ and $0\le j<N-1$, choose $a_{j,*}\in\operatorname*{arg\,max}_{1\le b\le D_j}\sigma_{j,b}^x$ when $\eta_j=x$, and any $a_{j,*}\in\{1,\ldots,D_j\}$ when $\eta_j=p$. Choose the corresponding vector $|e_j\rangle$ with $P_{j,a_{j,*}}|e_j\rangle=|e_j\rangle$ and $\langle e_j|e_j\rangle=1$.

The second moments $\sigma_{j,b}^{x}$ in Eq.~\eqref{eq:local-cost} already include the sum over all local control branches. Thus $a_{j,*}$ is selected after this branch average and does not depend on the branch subsequently sampled. No common maximizer of the individual branches is assumed. At the terminal time, choose
\[
a_*\in\operatorname*{arg\,max}_{1\le b\le D_{N-1}}
\frac{\sigma_{N-1,b}^{x}+\sigma_{N-1,\bar b}^{x}}2.
\]
Choose one normalized vector $|b\rangle$ from each eigenspace of $A_{N-1}$, so that $P_{N-1,b}|b\rangle=|b\rangle$ and $\langle b|c\rangle=\delta_{bc}$.
If $\lambda_{N-1,a_*}\ne0$, take $|\Psi_{a_*}\rangle=(|a_*\rangle+|\overline{a_*}\rangle)/\sqrt2$; if $\lambda_{N-1,a_*}=0$, take $|\Psi_{a_*}\rangle=|a_*\rangle$.

Using the same basis extension as above, choose $\rho_0=|e_0\rangle\langle e_0|$, take $U_{j+1:j}|e_j\rangle=|e_{j+1}\rangle$ for $0\le j<N-2$, and set
\[
U_{N-1:N-2}|e_{N-2}\rangle=|\Psi_{a_*}\rangle.
\]
This choice may depend on the fixed probabilities and estimators, but it is the same for every control sequence and every readout. At each measurement before the last, both the spectral phase control and the measurement operator act on the chosen eigenvector by multiplication by a scalar. More explicitly, for $0\le j<N-1$, every allowed branch, and every readout,
\[
T_{j,\beta}^{(\alpha),\eta_j}K_{j,\gamma}^{\eta_j}(y)|e_j\rangle\langle e_j|K_{j,\gamma}^{\eta_j}(y)^\dagger\bigl(T_{j,\beta}^{(\alpha),\eta_j}\bigr)^\dagger=k_{j,\gamma}^{\eta_j}(y;a_{j,*},a_{j,*})|e_j\rangle\langle e_j|.
\]
Consequently, the normalized state immediately before the terminal readout is $|\Psi_{a_*}\rangle\langle\Psi_{a_*}|$ for every preceding control sequence and readout vector with nonzero joint probability density. For $N=1$, choose this state directly as $\rho_0$. In both cases, with empty products set to one, the joint probability density of the complete control sequence and readout vector is
\[
\operatorname{Tr}\!\left[\rho_0 E_{\gamma,\mathbf U}^{\boldsymbol\eta,\mathrm{ctrl}}\bigl(\mathbf y,\mathbf T_{\boldsymbol\beta}^{(\boldsymbol\alpha),\boldsymbol\eta}\bigr)\right]=\prod_{j=0}^{N-2}\!\left[q_{j,\beta_j}^{(\alpha_j)}k_{j,\gamma}^{\eta_j}(y_j;a_{j,*},a_{j,*})\right]\frac{\phi_{\lambda_{N-1,a_*},\gamma}(y_{N-1})+\phi_{\lambda_{N-1,\overline{a_*}},\gamma}(y_{N-1})}{2}.
\]

The terminal control has only the branch $(\alpha_{N-1},\beta_{N-1})=(1,1)$ with probability one. The original terminal estimator satisfies, for $1\le b\le D_{N-1}$,
\[
\int h_{N-1,1}^{(1),x}(y,\gamma)
\phi_{\lambda_{N-1,b},\gamma}(y)\,dy
=\lambda_{N-1,b}.
\]
Summing over all control sequences and integrating all readouts in the preceding joint density therefore gives
\[
\mathrm E_{\rho_0,\mathbf U}f_{\mathbf T_{\boldsymbol\beta}^{(\boldsymbol\alpha),\boldsymbol\eta}}=\prod_{j=0}^{N-2}\!\left[\sum_{\alpha,\beta}q_{j,\beta}^{(\alpha)}\int h_{j,\beta}^{(\alpha),\eta_j}(y,\gamma)k_{j,\gamma}^{\eta_j}(y;a_{j,*},a_{j,*})\,dy\right]\frac{\lambda_{N-1,a_*}+\lambda_{N-1,\overline{a_*}}}2=0.
\]
The probabilities remain inside the branch sums throughout this calculation. The estimator is unchanged, and the same $\rho_0,\mathbf U$ is used in every term. Applying the identical sum and integral to the squared absolute value gives
\[
\mathrm E_{\rho_0,\mathbf U}\left|f_{\mathbf T_{\boldsymbol\beta}^{(\boldsymbol\alpha),\boldsymbol\eta}}\right|^2=\prod_{j=0}^{N-2}\!\left[\sum_{\alpha,\beta}q_{j,\beta}^{(\alpha)}\int\left|h_{j,\beta}^{(\alpha),\eta_j}(y,\gamma)\right|^2k_{j,\gamma}^{\eta_j}(y;a_{j,*},a_{j,*})\,dy\right]\frac{\sigma_{N-1,a_*}^x+\sigma_{N-1,\overline{a_*}}^x}{2}.
\]
With these eigenvectors and the choice of $a_*$ at the final measurement, this reduces to
\[
\mathrm E_{\rho_0,\mathbf U}\left|f_{\mathbf T_{\boldsymbol\beta}^{(\boldsymbol\alpha),\boldsymbol\eta}}\right|^2=\left(\prod_{j=0}^{N-2}\sigma_j^{\eta_j}\right)\max_b\frac{\sigma_{N-1,b}^x+\sigma_{N-1,\bar b}^x}{2}.
\]
The mean vanishes for this choice, so its variance equals its second moment. Maximization over all initial states and free evolutions can only increase that value. Hence every fixed feasible choice of probabilities and estimators obeys
\[
\max_{\rho_0,\mathbf U}\operatorname{Var}
f_{\mathbf T_{\boldsymbol\beta}^{(\boldsymbol\alpha),\boldsymbol\eta}}
\ge\left(\prod_{j=0}^{N-2}\sigma_j^{\eta_j}\right)
\max_b\frac{\sigma_{N-1,b}^x+\sigma_{N-1,\bar b}^x}{2}.
\]

To bound the maximum over $b$ in this expression from below by $\sigma_{N-1,*}^{x}(\gamma)$, consider the odd part of the final estimator,
\[
h_{N-1,1,\mathrm o}^{(1),x}(y,\gamma)
=\tfrac12\left[h_{N-1,1}^{(1),x}(y,\gamma)
-h_{N-1,1}^{(1),x}(-y,\gamma)\right].
\]
The identity $\phi_{\lambda,\gamma}(-y)=\phi_{-\lambda,\gamma}(y)$ and the calibration constraints on the final estimator give
\[
\int h_{N-1,1,\mathrm o}^{(1),x}(y,\gamma)\phi_{\lambda_{N-1,b},\gamma}(y)\,dy=\tfrac12(\lambda_{N-1,b}-\lambda_{N-1,\bar b})=\lambda_{N-1,b}.
\]

To bound its second moment, write $\sigma_{N-1,b}^x[\cdot]$ for the quantity in Eq.~\eqref{eq:local-cost} evaluated with the indicated terminal estimator. The odd part satisfies
\[
\left|h_{N-1,1,\mathrm o}^{(1),x}(y,\gamma)\right|^2\le\tfrac12\left(\left|h_{N-1,1}^{(1),x}(y,\gamma)\right|^2+\left|h_{N-1,1}^{(1),x}(-y,\gamma)\right|^2\right).
\]
Integrating this bound and using $\phi_{\lambda,\gamma}(-y)=\phi_{-\lambda,\gamma}(y)$ gives
\[
\sigma_{N-1,*}^x(\gamma)\le\max_b\sigma_{N-1,b}^x[h_{N-1,1,\mathrm o}^{(1),x}]\le\max_b\tfrac12\left(\sigma_{N-1,b}^x[h_{N-1,1}^{(1),x}]+\sigma_{N-1,\bar b}^x[h_{N-1,1}^{(1),x}]\right).
\]
This comparison applies to every original terminal estimator, including complex-valued ones. It does not restrict the optimization to odd functions or replace the estimator in the preceding probability calculation. Combining it with the variance lower bound yields, for every feasible choice of all local probabilities and estimators,
\[
\max_{\rho_0,\mathbf U}\operatorname{Var}
f_{\mathbf T_{\boldsymbol\beta}^{(\boldsymbol\alpha),\boldsymbol\eta}}
\ge\left(\prod_{j=0}^{N-2}\sigma_j^{\eta_j}\right)
\sigma_{N-1,*}^x(\gamma)
\ge\prod_{j=0}^{N-1}\sigma_{j,*}^{\eta_j}(\gamma).
\]

Only now minimize over the probabilities and estimators. The preceding inequality supplies a uniform lower bound. Equation~\eqref{eq:worst-bound}, together with independent local minimization, supplies the matching upper bound:
\[
\min_{\mathbf q,h}\max_{\rho_0,\mathbf U}\operatorname{Var}f_{\mathbf T_{\boldsymbol\beta}^{(\boldsymbol\alpha),\boldsymbol\eta}}\le\min_{\mathbf q,h}\prod_{j=0}^{N-1}\sigma_j^{\eta_j}=\prod_{j=0}^{N-1}\sigma_{j,*}^{\eta_j}(\gamma).
\]
The two bounds coincide, proving Eq.~\eqref{eq:exact-worst}.

At fixed $\gamma$, each factor can therefore be minimized independently. Minimizing their product over the common measurement strength gives the minimum worst-case variance stated in the main text, with $\gamma=\infty$ understood as the strong-measurement limit.

\subsection{Optimization problem}

For the two-time example with $J=3/2$ and $\boldsymbol\eta=(x,x)$, both measured observables have spectrum $(-1,-1/3,1/3,1)$. At fixed finite $\gamma>0$, Eq.~\eqref{eq:exact-worst} gives
\begin{equation}\label{eq:fixed-strength-problem}
\begin{aligned}
&\min_{\mathbf q,h}\max_{\rho_0,\mathbf U}\operatorname{Var}f_{\mathbf T_{\boldsymbol\beta}^{(\boldsymbol\alpha),(x,x)}} =\sigma_{0,*}^{x}(\gamma)\sigma_{1,*}^{x}(\gamma),\\
&\sigma_{j,*}^{x}(\gamma)=\min_{\mathbf q_j,h_j}\max_{\substack{\pi_{j,a}\ge0\\\sum_a\pi_{j,a}=1}}\sum_{a=1}^{4}\pi_{j,a}\sigma_{j,a}^{x}.
\end{aligned}
\end{equation}
The second moments $\sigma_{j,a}^{x}$ are defined in Eq.~\eqref{eq:local-cost}. The maximization over $\pi_{j,a}$ selects the largest of these second moments for each fixed choice of control probabilities and estimators.

At $j=0$, the branches are $(\alpha,\beta)=(1,1),(2,1),\ldots,(2,4)$, with $\upsilon_{0,\beta a}^{(\alpha)}=1-2\delta_{\alpha,2}\delta_{\beta,a}$. The probabilities satisfy $q_{0,\beta}^{(\alpha)}\ge0$ and $\sum_{\alpha,\beta}q_{0,\beta}^{(\alpha)}=1$. The calibration equations are
\begin{equation}\label{eq:opt-primal}
\sum_{\alpha,\beta}q_{0,\beta}^{(\alpha)}\upsilon_{0,\beta a}^{(\alpha)}\upsilon_{0,\beta b}^{(\alpha)}\int h_{0,\beta}^{(\alpha),x}(y,\gamma)k_{0,\gamma}^{x}(y;a,b)\,dy=\chi_{0,ab}^{x}.
\end{equation}

At $j=1$, the only branch has probability $q_{1,1}^{(1)}=1$, and the estimator must satisfy
\begin{equation}\label{eq:terminal-constraints}
\int h_{1,1}^{(1),x}(y,\gamma)\phi_{\lambda_{1,a},\gamma}(y)\,dy=\lambda_{1,a}.
\end{equation}
The calibration equations hold for every $1\le a\le b\le4$ at $j=0$ and every $1\le a\le4$ at $j=1$.

After solving both local problems at fixed strength, choose
\begin{equation}\label{eq:strength-selection}
\gamma_*\in\operatorname*{arg\,min}_{\gamma\in(0,\infty]}\sigma_{0,*}^{x}(\gamma)\sigma_{1,*}^{x}(\gamma).
\end{equation}
The endpoint $\gamma=\infty$ denotes the strong-measurement limit.

\subsection{Explicit optimality equations}

Equation~\eqref{eq:local-cost} allows complex-valued estimators. In this example, the calibration kernels, control signs, and target coefficients are real. Replacing each local estimator by its real part therefore preserves the constraints and cannot increase any $\sigma_{j,a}^{x}$. The local optimization in Eq.~\eqref{eq:fixed-strength-problem} can thus be restricted to real-valued estimators, for which the second-moment functionals are
\[
\begin{aligned}
&\sigma_{j,a}^{x}[\mathbf q_j,h_j;\gamma]=\sum_{\alpha,\beta}q_{j,\beta}^{(\alpha)}\int_{\mathbb R}[h_{j,\beta}^{(\alpha),x}(y,\gamma)]^2\phi_{\lambda_{j,a},\gamma}(y)\,dy,\\
&\sigma_j^{x}[\mathbf q_j,h_j;\gamma]=\max_{1\le a\le4}\sigma_{j,a}^{x}[\mathbf q_j,h_j;\gamma].
\end{aligned}
\]
The minimization remains subject to Eq.~\eqref{eq:opt-primal} for $j=0$ and Eq.~\eqref{eq:terminal-constraints} for $j=1$.

We use the Karush--Kuhn--Tucker (KKT) conditions~\cite{boyd2004} to solve the local optimization problems. Let $2z_{0,ab}$, $z_q$, and $\Lambda_{0,\beta}^{(\alpha)}\ge0$ be the multipliers for calibration, probability normalization, and $-q_{0,\beta}^{(\alpha)}\le0$, respectively. The Lagrangian for the first measurement is
\begin{equation}\label{eq:opt-loss}
\begin{aligned}
&\mathcal L_0^{\mathrm{loss}}={}\sum_a\pi_{0,a}\sigma_{0,a}^{x}+z_q\left(\sum_{\alpha,\beta}q_{0,\beta}^{(\alpha)}-1\right)-\sum_{\alpha,\beta}\Lambda_{0,\beta}^{(\alpha)}q_{0,\beta}^{(\alpha)}\\
&+2\sum_{a\le b}z_{0,ab}\left[\chi_{0,ab}^{x}-\sum_{\alpha,\beta}q_{0,\beta}^{(\alpha)}\upsilon_{0,\beta a}^{(\alpha)}\upsilon_{0,\beta b}^{(\alpha)}\int h_{0,\beta}^{(\alpha),x}(y,\gamma)k_{0,\gamma}^{x}(y;a,b)\,dy\right].
\end{aligned}
\end{equation}
Stationarity with respect to the estimators and branch probabilities, together with the calibration equations, gives
\begin{equation}\label{eq:nonfinal-finite-system}
\begin{aligned}
&h_{0,\beta}^{(\alpha),x}(y,\gamma)\sum_c\pi_{0,c}\phi_{\lambda_{0,c},\gamma}(y)
-\sum_{c\le d}z_{0,cd}\upsilon_{0,\beta c}^{(\alpha)}\upsilon_{0,\beta d}^{(\alpha)}k_{0,\gamma}^{x}(y;c,d)=0,\\
&z_q+\sum_c\pi_{0,c}\int [h_{0,\beta}^{(\alpha),x}(y,\gamma)]^2\phi_{\lambda_{0,c},\gamma}(y)\,dy\\
&-2\sum_{c\le d}z_{0,cd}\upsilon_{0,\beta c}^{(\alpha)}\upsilon_{0,\beta d}^{(\alpha)}\int h_{0,\beta}^{(\alpha),x}(y,\gamma)k_{0,\gamma}^{x}(y;c,d)\,dy-\Lambda_{0,\beta}^{(\alpha)}=0,\\
&\sum_{\alpha,\beta}q_{0,\beta}^{(\alpha)}\upsilon_{0,\beta a}^{(\alpha)}\upsilon_{0,\beta b}^{(\alpha)}\int h_{0,\beta}^{(\alpha),x}(y,\gamma)k_{0,\gamma}^{x}(y;a,b)\,dy=\chi_{0,ab}^{x}.
\end{aligned}
\end{equation}
The control probabilities satisfy $q_{0,\beta}^{(\alpha)}\ge0$ and $\sum_{\alpha,\beta}q_{0,\beta}^{(\alpha)}=1$. Each branch multiplier satisfies $\Lambda_{0,\beta}^{(\alpha)}\ge0$ and $q_{0,\beta}^{(\alpha)}\Lambda_{0,\beta}^{(\alpha)}=0$. The spectral weights satisfy $\pi_{0,a}\ge0$ and $\sum_a\pi_{0,a}=1$. For every spectral index, $\sigma_{0,a}^{x}\le\sigma_0^x$ and $\pi_{0,a}(\sigma_0^x-\sigma_{0,a}^{x})=0$. The calibration equation holds for each pair of spectral indices $(a,b)$ with $a\le b$. On a branch with zero probability, the estimator does not affect the protocol; we choose its otherwise arbitrary value to satisfy the stationarity condition for that estimator. With this convention, the system is equivalent to the KKT conditions for the convex formulation of the local optimization problem.

For $j=1$, let $2z_{1,aa}$ be the multipliers for the terminal calibration equations in Eq.~\eqref{eq:terminal-constraints}. The stationarity and calibration equations are
\begin{equation}\label{eq:terminal-finite-system}
\begin{aligned}
&h_{1,1}^{(1),x}(y,\gamma)\sum_c\pi_{1,c}\phi_{\lambda_{1,c},\gamma}(y)-\sum_c z_{1,cc}\phi_{\lambda_{1,c},\gamma}(y)=0,\\
&\int h_{1,1}^{(1),x}(y,\gamma)\phi_{\lambda_{1,a},\gamma}(y)\,dy=\lambda_{1,a}.
\end{aligned}
\end{equation}
Here $q_{1,1}^{(1)}=1$. The weights satisfy $\pi_{1,a}\ge0$ and $\sum_a\pi_{1,a}=1$; for each $a$, $\sigma_{1,a}^{x}\le\sigma_1^x$ and $\pi_{1,a}(\sigma_1^x-\sigma_{1,a}^{x})=0$. By definition, $\sigma_j^x$ is the largest of the second moments $\sigma_{j,a}^x$ in Eq.~\eqref{eq:local-cost}. Solving Eqs.~\eqref{eq:nonfinal-finite-system} and \eqref{eq:terminal-finite-system} with the feasibility and complementary slackness conditions stated after each gives the local minimizers in Eq.~\eqref{eq:fixed-strength-problem}.

\subsection{Solution for the collective-spin example}

For the spectra $(-1,-1/3,1/3,1)$, let $\bar a=5-a$ and let $\bar\beta=1$ for $\alpha=1$ and $\bar\beta=5-\beta$ for $\alpha=2$. Since $\lambda_{j,\bar a}=-\lambda_{j,a}$, the Gaussian density and the position-readout kernel satisfy
\begin{equation}\label{eq:example-reflection-data}
\begin{aligned}
&\phi_{\lambda_{j,\bar a},\gamma}(y)=\phi_{\lambda_{j,a},\gamma}(-y),\\
&k_{j,\gamma}^x(y;\bar b,\bar a)=k_{j,\gamma}^x(-y;a,b).
\end{aligned}
\end{equation}
The coefficients satisfy $\chi_{0,\bar b\bar a}^x=-\chi_{0,ab}^x$ and $\upsilon_{0,\bar\beta\bar a}^{(\alpha)}=\upsilon_{0,\beta a}^{(\alpha)}$. It follows that Eqs.~\eqref{eq:nonfinal-finite-system} and \eqref{eq:terminal-finite-system} are invariant under the following reflection. The estimators transform as
\begin{equation}\label{eq:example-reflection-transformation}
\begin{aligned}
&\widetilde h_{0,\beta}^{(\alpha),x}(y,\gamma)=-h_{0,\bar\beta}^{(\alpha),x}(-y,\gamma),\\
&\widetilde h_{1,1}^{(1),x}(y,\gamma)=-h_{1,1}^{(1),x}(-y,\gamma).
\end{aligned}
\end{equation}
The branch probabilities become $\widetilde q_{0,\beta}^{(\alpha)}=q_{0,\bar\beta}^{(\alpha)}$ and $\widetilde q_{1,1}^{(1)}=q_{1,1}^{(1)}=1$. The corresponding multipliers transform as $\widetilde\pi_{j,a}=\pi_{j,\bar a}$, $\widetilde z_{0,ab}=-z_{0,\bar b\bar a}$, $\widetilde z_{1,aa}=-z_{1,\bar a\bar a}$, and $\widetilde\Lambda_{0,\beta}^{(\alpha)}=\Lambda_{0,\bar\beta}^{(\alpha)}$, with $\widetilde z_q=z_q$ and $\widetilde\gamma=\gamma$.

Fix $\gamma$ and let $(\mathbf q_j,h_j)$ attain the minimum for time $j$ in Eq.~\eqref{eq:fixed-strength-problem}, so that $\sigma_j^x=\sigma_{j,*}^x(\gamma)$. Its reflected protocol satisfies the same calibration constraints and has second moments $\widetilde\sigma_{j,a}^x=\sigma_{j,\bar a}^x$. Average the branch probabilities and their products with the estimators:
\begin{equation}\label{eq:example-reflection-average}
\begin{aligned}
&\widehat q_{j,\beta}^{(\alpha)}=\frac{q_{j,\beta}^{(\alpha)}+\widetilde q_{j,\beta}^{(\alpha)}}2,\\
&\widehat q_{j,\beta}^{(\alpha)}\widehat h_{j,\beta}^{(\alpha),x}(y,\gamma) =\frac{q_{j,\beta}^{(\alpha)}h_{j,\beta}^{(\alpha),x}(y,\gamma)+\widetilde q_{j,\beta}^{(\alpha)}\widetilde h_{j,\beta}^{(\alpha),x}(y,\gamma)}2.
\end{aligned}
\end{equation}
If $\widehat q_{j,\beta}^{(\alpha)}=0$, the corresponding estimator $\widehat h_{j,\beta}^{(\alpha),x}(y,\gamma)$ may be set to zero. Reflection interchanges the two terms in each average, so the averaged protocol is invariant under reflection. The averaged probabilities remain nonnegative and normalized. Each calibration equation is linear in the products $qh$, and both original and reflected protocols give its required target value. Their average therefore satisfies the same calibration equation. Hence the averaged protocol is feasible.

Write $\widehat\sigma_{j,a}^x\equiv\sigma_{j,a}^x[\widehat{\mathbf q}_j,\widehat h_j;\gamma]$ for its second moments and $\widehat\sigma_j^x\equiv\max_a\widehat\sigma_{j,a}^x$ for their maximum. Since $\sigma_{j,*}^x(\gamma)$ is the minimum over all feasible protocols, feasibility gives
\[
\sigma_{j,*}^x(\gamma)\le\widehat\sigma_j^x.
\]
On the other hand, applying the Cauchy--Schwarz inequality to the two terms on the right-hand side of the second line of Eq.~\eqref{eq:example-reflection-average}, integrating with the nonnegative density $\phi_{\lambda_{j,a},\gamma}(y)$, and summing over the control branches gives
\begin{equation}\label{eq:example-reflection-cost}
\widehat\sigma_{j,a}^x\le\frac{\sigma_{j,a}^x+\sigma_{j,\bar a}^x}{2}\le\sigma_j^x=\sigma_{j,*}^x(\gamma).
\end{equation}
Taking the maximum over $a$ and combining the two bounds yields
\[
\sigma_{j,*}^x(\gamma)\le\widehat\sigma_j^x\le\sigma_{j,*}^x(\gamma).
\]
Thus $\widehat\sigma_j^x=\sigma_{j,*}^x(\gamma)$: symmetrizing any optimal protocol gives another optimal protocol. An optimal protocol can therefore be chosen invariant under reflection.

At this symmetric optimum, averaging the multipliers $\pi_{j,a},z_{0,ab},z_{1,aa},z_q$ with their reflected values preserves the KKT conditions on every positive-probability branch. For a zero-probability branch, the stationarity conditions for the estimator and the branch probability in Eq.~\eqref{eq:nonfinal-finite-system} require
\begin{equation}\label{eq:example-reflection-inactive}
\int\frac{\left[\sum_{c\le d}z_{0,cd}\upsilon_{0,\beta c}^{(\alpha)}\upsilon_{0,\beta d}^{(\alpha)}k_{0,\gamma}^x(y;c,d)\right]^2}{\sum_c\pi_{0,c}\phi_{\lambda_{0,c},\gamma}(y)}\,dy\le z_q.
\end{equation}
The integrand is jointly convex in the linear combination inside the square and the positive denominator, so averaging the multipliers preserves this inequality. The stationarity conditions in Eq.~\eqref{eq:nonfinal-finite-system} then determine the estimator and the nonnegative multiplier $\Lambda_{0,\beta}^{(\alpha)}$ on this branch. We can thus choose estimators satisfying
\begin{equation}\label{eq:example-reflection-fixedpoint}
\begin{aligned}
&h_{0,\beta}^{(\alpha),x}(y,\gamma)=-h_{0,\bar\beta}^{(\alpha),x}(-y,\gamma),\\
&h_{1,1}^{(1),x}(y,\gamma)=-h_{1,1}^{(1),x}(-y,\gamma).
\end{aligned}
\end{equation}
The corresponding probabilities obey $q_{0,\beta}^{(\alpha)}=q_{0,\bar\beta}^{(\alpha)}$, and the multipliers obey $\Lambda_{0,\beta}^{(\alpha)}=\Lambda_{0,\bar\beta}^{(\alpha)}$, $\pi_{j,a}=\pi_{j,\bar a}$, $z_{0,ab}=-z_{0,\bar b\bar a}$, and $z_{1,aa}=-z_{1,\bar a\bar a}$.

\textit{Symmetric solution.}
The stationarity conditions for the estimators in Eqs.~\eqref{eq:nonfinal-finite-system} and \eqref{eq:terminal-finite-system} give
\begin{equation}\label{eq:opt-estimator-explicit}
\begin{aligned}
&h_{0,\beta}^{(\alpha),x}(y,\gamma)=\frac{\sum_{c\le d}z_{0,cd}\upsilon_{0,\beta c}^{(\alpha)}\upsilon_{0,\beta d}^{(\alpha)}k_{0,\gamma}^{x}(y;c,d)}{\sum_c\pi_{0,c}\phi_{\lambda_{0,c},\gamma}(y)},\\
&h_{1,1}^{(1),x}(y,\gamma)=\frac{\sum_c z_{1,cc}\phi_{\lambda_{1,c},\gamma}(y)}{\sum_c\pi_{1,c}\phi_{\lambda_{1,c},\gamma}(y)}.
\end{aligned}
\end{equation}

For the solution used in the main text, set $q\equiv q_{0,1}^{(1)}$, with $0<q<1$. The probabilities of the four phase-flip branches and the spectral weights at the final measurement are
\begin{equation}\label{eq:spin-support}
\begin{aligned}
&\left(q_{0,1}^{(2)},q_{0,2}^{(2)},q_{0,3}^{(2)},q_{0,4}^{(2)}\right)=\frac{1-q}{2}(1,0,0,1),\\
&\left(\pi_{1,1},\pi_{1,2},\pi_{1,3},\pi_{1,4}\right)=\frac12(1,0,0,1).
\end{aligned}
\end{equation}
At the first measurement, $\pi_{0,1}=\pi_{0,4}>0$ and $\pi_{0,2}=\pi_{0,3}>0$. The zero probabilities in this choice are determined by the KKT conditions in addition to reflection symmetry.

At fixed $\gamma$, set $s=\gamma^2y/3$. Substitution of the symmetric multipliers into Eq.~\eqref{eq:opt-estimator-explicit} gives
\begin{equation}\label{eq:example-optimal-estimators}
\begin{aligned}
&h_{0,1}^{(1),x}(y,\gamma)=\frac{u_1\sinh3s+u_2\sinh2s+u_3\sinh s}{2\cosh s[\cosh(2s)+\zeta]},\\
&h_{0,1}^{(2),x}(y,\gamma)=\frac{u_1\sinh3s+u_2\cosh2s+u_3\sinh s+u_4e^{-s}}{2\cosh s[\cosh(2s)+\zeta]},\\
&h_{0,2}^{(2),x}(y,\gamma)=\frac{u_1\sinh3s+u_2\cosh2s+u_3\sinh s-u_4e^s}{2\cosh s[\cosh(2s)+\zeta]},\\
&h_{1,1}^{(1),x}(y,\gamma)=\frac{v_1\sinh3s+v_2\sinh s}{\cosh3s}.
\end{aligned}
\end{equation}
Reflection determines the estimators for the other two phase-flip branches, $h_{0,4}^{(2),x}(y,\gamma)=-h_{0,1}^{(2),x}(-y,\gamma)$ and $h_{0,3}^{(2),x}(y,\gamma)=-h_{0,2}^{(2),x}(-y,\gamma)$.
The coefficients at fixed $\gamma$ are determined by the multipliers:
\begin{equation}\label{eq:main-coefficients-forward}
\begin{aligned}
&u_2=-\frac{e^{2\gamma^2/9}z_{0,12}}{\pi_{0,1}},\\
&u_3=-\frac{e^{2\gamma^2/9}z_{0,13}+e^{4\gamma^2/9}z_{0,22}}{\pi_{0,1}},\\
&u_4=-\frac{e^{2\gamma^2/9}z_{0,13}}{\pi_{0,1}},\\
&\zeta=\frac12\left(\frac{\pi_{0,2}}{\pi_{0,1}}e^{4\gamma^2/9}-1\right).
\end{aligned}
\end{equation}
The remaining coefficients are $u_1=-z_{0,11}/\pi_{0,1}$, $v_1=-2z_{1,11}$, and $v_2=-2e^{4\gamma^2/9}z_{1,22}$.

\textit{Equations for the coefficients.}
Equation~\eqref{eq:main-coefficients-forward} and normalization express the spectral weights in terms of $\zeta$:
\begin{equation}\label{eq:spin-pi-coefficients}
\begin{aligned}
&\pi_{0,1}=\pi_{0,4}=\frac{1}{2[1+(2\zeta+1)e^{-4\gamma^2/9}]},\\
&\pi_{0,2}=\pi_{0,3}=\frac{(2\zeta+1)e^{-4\gamma^2/9}}{2[1+(2\zeta+1)e^{-4\gamma^2/9}]}.
\end{aligned}
\end{equation}

At fixed $\gamma$, the unknowns at the first measurement are $q,u_1,u_2,u_3,u_4,\zeta,z_q,\sigma_0^x$. Substituting Eq.~\eqref{eq:example-optimal-estimators} into the calibration and KKT conditions, and removing their reflected duplicates, gives the following eight scalar equations:
\begingroup
\allowdisplaybreaks[1]
\begin{subequations}\label{eq:spin-coefficient-system}
\begin{align}
&\int\frac{u_1\sinh3s+q u_2\sinh2s+[u_3-(1-q)u_4]\sinh s}{2\cosh s[\cosh(2s)+\zeta]}\phi_{-1,\gamma}(y)\,dy=-1,\label{eq:spin-diagonal-outer}\\
&\int\frac{u_1\sinh3s+q u_2\sinh2s+[u_3-(1-q)u_4]\sinh s}{2\cosh s[\cosh(2s)+\zeta]}\phi_{-1/3,\gamma}(y)\,dy=-\frac13,\label{eq:spin-diagonal-inner}\\
&q e^{-\gamma^2/18}\int\frac{u_1\sinh3s+u_2\sinh2s+u_3\sinh s}{2\cosh s[\cosh(2s)+\zeta]}\phi_{-2/3,\gamma}(y)\,dy\notag\\*
&-(1-q)e^{-\gamma^2/18}\int\frac{u_2\cosh2s+u_4\cosh s}{2\cosh s[\cosh(2s)+\zeta]}\phi_{-2/3,\gamma}(y)\,dy=-\frac23,\label{eq:spin-calibration-12}\\
&q e^{-2\gamma^2/9}\int\frac{u_1\sinh3s+u_2\sinh2s+u_3\sinh s}{2\cosh s[\cosh(2s)+\zeta]}\phi_{-1/3,\gamma}(y)\,dy\notag\\*
&-(1-q)e^{-2\gamma^2/9}\int\frac{u_2\cosh2s+u_4\cosh s}{2\cosh s[\cosh(2s)+\zeta]}\phi_{-1/3,\gamma}(y)\,dy=-\frac13,\label{eq:spin-calibration-13}\\
&z_q=\sum_{c=1}^{4}\pi_{0,c}\int\left[\frac{u_1\sinh3s+u_2\sinh2s+u_3\sinh s}{2\cosh s[\cosh(2s)+\zeta]}\right]^2\phi_{\lambda_{0,c},\gamma}(y)\,dy,\label{eq:spin-probability-identity}\\
&z_q=\sum_{c=1}^{4}\pi_{0,c}\int\left[\frac{u_1\sinh3s+u_2\cosh2s+u_3\sinh s+u_4e^{-s}}{2\cosh s[\cosh(2s)+\zeta]}\right]^2\phi_{\lambda_{0,c},\gamma}(y)\,dy,\label{eq:spin-probability-flip}\\
&\sigma_0^x=q\int\left[\frac{u_1\sinh3s+u_2\sinh2s+u_3\sinh s}{2\cosh s[\cosh(2s)+\zeta]}\right]^2\phi_{-1,\gamma}(y)\,dy\notag\\*
&+(1-q)\int\frac{[u_1\sinh3s+(u_3-u_4)\sinh s]^2+[u_2\cosh2s+u_4\cosh s]^2}{4\cosh^2s[\cosh(2s)+\zeta]^2}\phi_{-1,\gamma}(y)\,dy,\label{eq:spin-worst-outer}\\
&\sigma_0^x=q\int\left[\frac{u_1\sinh3s+u_2\sinh2s+u_3\sinh s}{2\cosh s[\cosh(2s)+\zeta]}\right]^2\phi_{-1/3,\gamma}(y)\,dy\notag\\*
&+(1-q)\int\frac{[u_1\sinh3s+(u_3-u_4)\sinh s]^2+[u_2\cosh2s+u_4\cosh s]^2}{4\cosh^2s[\cosh(2s)+\zeta]^2}\phi_{-1/3,\gamma}(y)\,dy.\label{eq:spin-worst-inner}
\end{align}
\end{subequations}
\endgroup

The coefficients $v_1,v_2$ at the final measurement satisfy
\begin{subequations}\label{eq:spin-terminal-coefficients}
\begin{align}
&v_1\int\frac{\sinh3s}{\cosh3s}\phi_{-1,\gamma}(y)\,dy+v_2\int\frac{\sinh s}{\cosh3s}\phi_{-1,\gamma}(y)\,dy=-1,\label{eq:spin-terminal-outer}\\
&v_1\int\frac{\sinh3s}{\cosh3s}\phi_{-1/3,\gamma}(y)\,dy+v_2\int\frac{\sinh s}{\cosh3s}\phi_{-1/3,\gamma}(y)\,dy=-\frac13.\label{eq:spin-terminal-inner}
\end{align}
\end{subequations}

The solutions must satisfy $0<q<1$, $\zeta>-1/2$, and the remaining KKT inequalities
\begin{subequations}\label{eq:spin-coefficient-inequalities}
\begin{align}
&z_q\ge\sum_{c=1}^{4}\pi_{0,c}\int\left[\frac{u_1\sinh3s+u_2\cosh2s+u_3\sinh s-u_4e^s}{2\cosh s[\cosh(2s)+\zeta]}\right]^2\phi_{\lambda_{0,c},\gamma}(y)\,dy,\label{eq:spin-unused-inequality}\\
&\int\left[\frac{v_1\sinh3s+v_2\sinh s}{\cosh3s}\right]^2\left[\phi_{-1,\gamma}(y)-\phi_{-1/3,\gamma}(y)\right]\,dy\ge0.\label{eq:spin-terminal-inequality}
\end{align}
\end{subequations}

A numerical program solves these systems subject to the inequalities and selects the common measurement strength according to Eq.~\eqref{eq:strength-selection}.
For the protocol used in the main text, the optimal strength is $\gamma_*\simeq2.16$ and the identity-branch probability is $q_{0,1,*}^{(1)}\simeq0.33$. The first-measurement coefficients are
\[
(u_1,u_2,u_3,u_4,\zeta)\simeq(1.40,0.91,-3.04,3.52,0.19).
\]
The final measurement gives $(v_1,v_2)\simeq(1.39,-5.25)$. All values are rounded to two decimal places.